\documentstyle[11pt]{article}
\def\up#1{\raise 1ex\hbox{#1}}
\begin{document}
\centerline{\bf STOCHASTIC DIFFERENTIAL GEOMETRY AND THE RANDOM}
\centerline{\bf FLOWS OF VISCOUS AND MAGNETIZED FLUIDS} 
\centerline{\bf IN SMOOTH MANIFOLDS AND EUCLIDEAN SPACE}

\bigskip
\centerline{by}
\bigskip
\centerline{\bf Diego L. Rapoport}
\bigskip
\centerline{Isaac Newton Institute Institute for Mathematical Sciences}
\centerline {20 Clarkson Rd., Cambridge, CB3 0EH, U.K.}
\centerline{Permanent address: Av. Alvarez Thomas 646, Buenos Aires 1427}
\centerline{Argentina; drapo@unq.edu.ar}

\bigskip

{\bf Abstract:} In this article we integrate in closed implicit form the Navier-Stoker equations
for an incompressible fluid in a smooth compact manifold without boundary, and
in particular, in a compact manifold which is isometrically
embedded in Euclidean space, and finally in Euclidean space itself. We further integrate the kinematic dynamo
problem of magnetohydrodynamics, i.e. the equations of passive transport
of a magnetic field on a fluid. We carry out these integrations through the
application of the methods of Stochastic Differential Geometry, i.e. the gauge theoretical formulation 
of diffusion processes on smooth manifolds. Thus we start by defining the invariant infinitesimal generators of diffusion processes
of differential forms on smooth compact manifolds, in terms of the laplacians (on differential forms)
associated with the Riemann-Cartan-Weyl (RCW) metric compatible connections. These
geometries have a torsion tensor which reduces to a trace $1$-form, whose
conjugate vector field is the drift of the diffusion of scalar fields. We construct the diffusion
processes of differential forms associated with these laplacians by
using the property that the solution flow of the stochastic
differential equation corresponding to the scalar diffusion is -assuming Hoelder
or Sobolev
regularity conditions- a random diffeomorphism
of the manifold. We apply these constructions to 
give a new characterization of the Navier-Stokes equation for the velocity one-form of an incompressible fluid as a non-linear diffusion process determined by 
a RCW connection. We prove the Navier-Stokes equations to be equivalent to a linear diffusion
equation for the vorticity and the Poisson-de Rham equation for the velocity
with the vorticity as a source, extending thus to manifolds, the approach in practice in the random vortex method in Computational Fluid Dynamics.  We give the invariant random stochastic differential
equations for the position (as a Lagrangian representation) of the fluid particles and
thus obtain a random diffeomorphism which is a solution of the Navier-Stokes equation. 
We solve the Cauchy problem for the heat equation for the vorticity two-form and the Dirichlet problem for the Poisson-de Rham equation
for the velocity one-form in two different instances. Firstly, using the gradient flows obtained by a Nash isometric embedding of the manifold in Euclidean space, and secondly, for the more general case of an arbitrary manifold, by running curvature and fluid-deformation tensor dependant random flows on the tangent bundle, projecting on the random fluid' particles flow. We implement our general construction on Euclidean space by simply taking the former construction under the identity embedding. We extend these methods to 
the integration of the kinematic dynamo problem. From the fact that any metric compatible connection can be constructed
in terms of the push-forward connection and using simple stochastic analysis considerations, we prove that any
diffusion process generated by a RCW connection admits a random flow representable through a purely diffusive process (i.e. zero drift), in any dimension other than $1$. We apply this to prove that the random flow of a viscous fluid obeying the classical Navier-Stokes equations on a smooth manifold of dimension other than $1$, can be represented as a purely diffusive process, where the new diffusion tensor can be constructed in terms of the velocity and the original diffusion tensor associated to the metric and the kinematical viscosity. A similar construction we prove to be valid for the fluid flow of a passively magnetized viscous or inviscid fluid.
\eject

\section{Introduction.}

\hspace{2em} 
The purpose of this article is two-fold: Firstly to give implicit random 
representations for the solutions of the Navier-Stokes equation for an incompressible
fluid and for the kinematic dynamo problem of magnetohydrodynamics, in several instances; firstly, on an arbitrary compact orientable smooth 
manifold (without boundary) -following our presentation in [58,59]- and further in the case in which the
manifold is isometrically embedded in
Euclidean space, to finally give the expressions for
Euclidean space itself. Secondly, to present as a basis for
such an integration, the gauge-theoretical structures of
Brownian motion theory and the stochastic analysis rules associated to them.
Thus, the method of integration we shall apply for our objectivs stems
from stochastic differential geometry, i.e. the gauge theory of Brownian processes in smooth
manifolds and Euclidean space developed in the pioneering works by Ito [15], Elworthy and Eells [13], P. Malliavin [11]
and further elaborated by Elworthy [12], Ikeda and Watanabe [14], P. Meyer [40], and Rogers and Williams [37].
Associated to these geometrical structures which can be
written in terms of the Cartan calculus on manifolds of classical differential
geometry,  we shall present the rules of stochastic analysis which 
describe the transformation of differential forms along 
the paths of generalized Brownian motion generated by these geometries, setting
thus the method for the integration of linear evolution 
equations for differential forms; this is the well known martingale problem approach
to the solution of partial differential equations on manifolds [30].

While classical Hamiltonean systems with finite degrees of freedom may appear to
have a random behavior, in fluid dynamics it is known that
the Euler equation for an inviscid fluid
is a hamiltonean system with infinite degrees of freedom supporting as well infinite
conserved quantities; such a system appears to be non-random [56]. The situation
is radically different in the case of a viscous fluid described by the Navier-Stokes equations.
In this case, there is a second-order partial derivative associated to the kinematical
viscosity, which points out to the fact that there is a diffusion term, which can be
thougth as related to a Brownian motion. Thus in the viscous case, there is from the very
beginning a random element. While in the Euler case the group of interest is the
(Riemannian) volume preserving of diffeomorphims, it will turn out in the course of
these studies, that there is an active group of ${\bf random}$ diffeomorphims which
represent the Lagrangian random flow of the viscous fluid´particles. In this case, when
there is a non-constant diffusion tensor describing the local amplification of noise,
these diffeomorphisms do not preserve the volume measure, contributing at a dynamical
level -as it will turn out-
to the complicated topology of turbulent and magnetized flows [56].
 
The essential role of randomness in Fluid Dynamics appears already at an experimental level. The analysis of the velocimetry signal of a turbulent fluid shows that its velocity
is a random variable, even though that the dynamics is ruled by the deterministic
Navier-Stokes equation [8].
The concept of a turbulent fluid as a stochastic process was first proposed
by Reynolds [16], who decomposed the velocity into mean velocity plus fluctuations.
The Reynolds approach is currently used in most numerical
simulations of turbulent fluids in spite of the fact that it leads to unsurmountable non-closure problems of the transport equations.; see Lumley [18], Mollo and Christiansen [38]. 
Furthermore, the Reynolds decomposition is non invariant 
alike the usual decomposition into drift and white noise perturbation
in the non-invariant theory of diffusion processes.
Other treatments of stochasticity in turbulence 
were advanced from the point of view of Feynman path integrals, as initiated by Monin and Yaglom [17].
From the point of view of
diffusion processes, invariant measures for stochastic modifications of the Navier-Stokes
equations on euclidean domains, have been constructed by Vishik and Fursikov [36]
and Cruzeiro and Albeverio [42]. (It is important to remark that
the existance of an invariant measure for NS as a ${\it classical}$ dynamical system
is the starting point of the classical dynamical systems approach to turbulence; see Ruelle [48].)
Contemporary investigations develop
the relations between randomness and the many-scale structure of turbulence 
which stems from the Kolmogorov theory as presented by Fritsch[8] and Lesieur [3],
and apply the renormalization group method; see Orzsag [19]. 

A completely new line of research followed from the understanding
of the fundamental importance of the vorticity (already stressed by Leonardo da Vinci)
in the self-organization of turbulent fluids, which was assessed by numerical simulations by
Lesieur [3,4], and theoretically by Majda [39] and Chorin [1].
It was found that 
the Navier-Stokes equations for an incompressible fluids on Euclidean domains
yields a linear diffusion equation for the vorticity
which becomes a source for the velocity through the Poisson equation: Solving the latter equation,
we can obtain an expression for the fluid´s velocity in 2D.
This observation was the starting point for the
random vortex method in Computational Fluid Mechanics largely due to Chorin [1,2,6]. This conception lead 
to apply methods of statistical mechanics (as originally proposed by Onsager [20]) to study the complex topology of vortex dynamics
and to relate this to polymer dynamics [1].
In the random vortex method a random lagrangian representation for the position
of the incompressible
fluid particles was proposed. Consequently, the Navier-Stokes linear (¨heat¨) equation for the vorticity
was integrated only for two dimensional fluids (implicit to this is the martingale problem approach quoted
above), while
the general case was numerically integrated by discretization of the
this ¨heat¨ equation; see Chorin [1,2]. The difficulty for the exact integration
in the general case apparently stems from the fact
that while in dimension
$2$ the vorticity $2$-form can be identified with a density and then the integration
of the Navier-Stokes equation for the vorticity follows from the application of the
well known Ito formula for scalar fields, in the case of higher dimension this identification
is no longer valid and an Ito formula for $2$-forms is required to carry out the
integration. This formula became only recently available in the works by Elworthy
[27] and Kunita [24], in the context of the theory of random flows
on smooth manifolds.

The importance of a Stochastic Differential Geometry treatment of the Navier-Stokes equation on a smooth
n-manifold $M$ stems from several fundamental facts which are keenly interwoven. For a start, it provides an intrinsic geometrical characterization
of diffusion processes of differential forms which follows from the characterizations of the laplacians associated to non-Riemannian geometries with torsion
of the trace type, as the infinitesimal generators of the diffusions. In particular, this will allow to obtain a new way
of writing the Navier-Stokes equation for an incompressible fluid
in terms of these laplacians acting on differential one-forms (velocities)
and two-forms (vorticities). Furthermore, 
these diffusion processes of differential forms, are constructed starting
from the scalar diffusion process which under Hoelder continuity or Sobolev
regularity conditions,
yields a time-dependant random diffeomorphism of $M$ which will
represent the Lagrangian trajectories for the fluid particles position. This diffeomorphic
property will allow us to use the Ito formula for differential forms (following the
presentation due to Elworthy)
as the key instrument for the integration of the Navier-Stokes equation (NS for short, in the following)
for an incompressible fluid. Thus, it is the knowledge of the rules of stochastic analysis
what sets the martingale problem approach to the solution of NS when he have transformed
it to an equivalent system which is essentially linear. This transformation of non-linear
equations to linear ones appears as well in an identical gauge-theoretical approach
to quantum mechanics as a theory of Brownian motion. In this case, the massive
Dirac-Hestenes equation for a Dirac-Hestenes operator field (conceived as a geometrical
matter field) can be transformed in 
a Clifford bundle approach to the sourceless Maxwell equation [26]. We must remark that
the geometrical structures on which the gauge-theoretical foundations of Brownian
motion are introduced, belong in fact originally to the gauge theory of gravitation, including not
only translational degrees of freedom, but additionally spinor fields [46,57]. 

For the benefit of completeness, we have included
in this article an Appendix which reviews briefly the main probabilistic and analytical
notions appearing in this article. This article represents the contents of a lecture delivered in December 18, 2000, at the Seminar in Geometry and Topology
of Fluid Flows, Isaac Newton Institute for Mathematical Sciences, and expands on
previous results [59,62,63]. 

\section{Riemann-Cartan-Weyl Geometry of Diffusions}
\hspace{2em}
The objective of this section is to show that the invariant definition of a "heat"
(Fokker-Planck) operator requires the introduction of linear connections of a certain
type. 

Let us consider for a start, a smooth $n$-dimensional compact orientable manifold $M$ (without boundary), on
which we shall consider a second-order smooth differential operator $L$.
On a local coordinate system, $(x^\alpha), \alpha =1, \ldots ,n$, $L$ is written
as
\begin{eqnarray}
L = {1 \over 2}g^{\alpha \beta}(x)\partial_\alpha \partial _\beta +
B^\alpha (x) \partial _\alpha + c(x).
\end{eqnarray}
From now on, we shall fix this coordinate system, and all local
expressions shall be written in it.

Although formally, there is no restriction as to the nature of
$M$, we are really thinking on a n-dimensional space (or space-time)
manifold, and {\bf not} in a phase-space
manifold of a dynamical system. 

The {\it principal symbol} $\sigma$ of $L$, is the section of the bundle
of real bilinear symmetric maps on $T^*M$, defined as follows: for $x \in
M$, $p_i \in T^*_xM$, take $C^2$ functions, $f_i:M \to R$ with $f_i(x) =
0$ and $df_i(x) = p_i$, $i=1,2$; then,
\begin{eqnarray}
\sigma (x)(p_1,p_2) = L(f_1f_2)(x). 
\end{eqnarray}
Note that $\sigma$ is well defined, i.e., it is independent of the choice
of the functions $f_i, i=1,2$.

If $L$ is locally as in $(1)$, then $\sigma$ is locally represented by the
matrix $(g^{\alpha \beta})$. We can also view $\sigma$ as a section of the
bundle of linear maps $L(T^*M,TM)$, or as a section of the bundle $TM
\otimes TM$, or still as a bundle morphism from $T^*M$ to $TM$. If
$\sigma$ is a bundle isomorphism, it induces a Riemannian structure
$g$ on $M$, $g:M \to L(TM,TM)$:
\begin{eqnarray*}
g(x)(v_1,v_2):= <\sigma(x)^{-1}v_1,v_2>_x,
\end{eqnarray*}
for $x \in M$, $v_1, v_2 \in T^*_xM$. Here, $<.,.>_x$ denotes the natural duality
between $T^*_xM$ and $T_xM$. Locally, $g(x)$ is represented by the matrix
${1 \over 2}(g^{\alpha \beta}(x))^{-1}$. Consider the quadratic forms over
$M$ associated to $L$, defined as
\begin{eqnarray*}
Q_x(p_x) = {1 \over 2}<p_x,\sigma _x(p_x)>_x,
\end{eqnarray*}
for $x \in M$, $p_x \in T^*_xM$. Then, with the local representation $(1)$
for $L$, $Q_x$ is represented as ${1 \over 2}(g^{\alpha \beta}(x))$. Then,
$L$ is an elliptic (semi-elliptic) operator whenever for all $x \in M$,
$Q_x$ is positive-definite (non-negative definite). We shall assume in the
following that $L$ is an elliptic operator. In this case, $\sigma$ is a bundle
isomorphism and the metric $g$ is actually a Riemannian metric. Notice, as
well, that $\sigma (df) = grad~f$, for any $f:M \to R$ of class $C^2$,
where ${\rm grad}$ denotes the Riemannian gradient.

We wish to give an intrinsic description of $L$, i.e. a description
independent of the local coordinate system. This is the essential prerequisite
of covariance.

For this, we shall introduce for the general setting of the discussion,
an arbitrary connection on $M$, whose covariant
derivative we shall denote as $\tilde\nabla$. We remark here that $\tilde\nabla$ will ${\bf not}$ be the Levi-Civita connection associated to $g$; we shall
precise this below. Let $\sigma (\tilde\nabla)$ denote the second-order
part of $L$, and let us denote by $X_0(\tilde\nabla)$ the vector field on $M$
given by the first-order part of $L$. Finally, the zero-th order part of $L$
is given by $L(1)$, where $1$ denotes the constant function on $M$ equal
to $1$. We shall assume in the following, that $L(1)$ vanishes identically.

Then, for $f:M \to R$ of class $C^2$, we have
\begin{eqnarray}
\sigma (\tilde\nabla)(x) = {1 \over 2}trace (\tilde\nabla^2f)(x) = {1 \over 2}(\tilde\nabla df)(x)),
\end{eqnarray}
where the trace is taken in terms of $g$, and $\tilde\nabla df$ is thought as a
section of $L(T^*M,T^*M)$. Also, $X_0(\tilde\nabla) =  L - \sigma
(\tilde\nabla)$. If $\tilde\Gamma ^\alpha _{\beta \gamma}$ is the local representation
for the Christoffel symbols of the connection, i.e. $\tilde\nabla_{\partial \over \partial x^ \alpha}{\partial \over \partial x^ \beta}
=  \tilde\Gamma^\nu_{\alpha \beta}{\partial \over \partial x^ \nu}$, then the local
representation of $\sigma (\tilde\nabla)$ is:
\begin{eqnarray}
\sigma (\tilde\nabla)(x) = {1 \over 2}g^{\alpha \beta}(x)(\partial _\alpha
\partial _\beta + \tilde\Gamma ^\gamma _{\alpha \beta}(x)\partial
_\gamma),
\end{eqnarray}
and
\begin{eqnarray}
X_0(\tilde\nabla)(x) = B^\alpha(x)\partial _\alpha - {1 \over 2}g^{\alpha
\beta}(x)\tilde\Gamma ^\gamma _{\alpha \beta}\partial _\gamma. 
\end{eqnarray}
If in particular, we take $\tilde\nabla$ the Levi-Civita connection associated to $g$, which we
shall denote as $\nabla ^g$, then for any $f:M \to R$ of class $C^2$:
\begin{eqnarray}
\sigma (\nabla ^g)(df) = {1 \over 2}{\rm trace}((\nabla^g)^2f) = {1 \over 2}{\rm trace}(\nabla ^g df) = -{1/2}
{\rm div_g~ grad}~f = {1/2}\triangle_g f. 
\end{eqnarray}
Here, $\triangle _g$ is the Levi-Civita laplacian operator on functions; locally it
is written as
\begin{eqnarray}
\triangle _g =  g^{-1/2}\partial _\alpha ((g^
{1/2}g^{\alpha \beta}\partial _\beta);~~ g = det(g_{\alpha \beta}),
\end{eqnarray}
and $div_g$ is the Riemannian divergence operator on vector fields $X = X^{\alpha}(x)\partial_\alpha$:
\begin{eqnarray}
div_g(X) = -g^{-1/2}\partial_{\alpha}(g^{1/2}X^{\alpha}). 
\end{eqnarray}
Note the relation we already have used in eqt. $(6)$ and will be used repeatedly; namely:
\begin{eqnarray}
div_g(X) = -\delta \tilde X,
\end{eqnarray}
where $\delta$ is the co-differential operator (see $(23)$ below), and $\tilde X$ is the one-form conjugate to the vector field $X$, i.e. $\tilde
X_\alpha = g_{\alpha \beta}X^\beta$.

We now take $\tilde\nabla$ to be a Cartan connection with torsion [10,46], which we additionally
assume to be compatible with $g$, i.e. $\tilde\nabla g = 0$. Then $\sigma
(\tilde\nabla) = {1 \over 2}{\rm trace}(\tilde\nabla ^2)$. Let us compute this. Denote the Christoffel
coefficients of $\tilde\nabla$ as $\tilde\Gamma^\alpha _{\beta \gamma}$; then,
\begin{eqnarray}
\tilde\Gamma^\alpha_{\beta\gamma} = {\alpha\brace\beta\gamma}~ +
1/2K^\alpha_{\beta\gamma},
\end{eqnarray}
where the first term in $(10)$ stands for the Christoffel Levi-Civita coefficients of the
metric $g$, and 
\begin{eqnarray}
K^\alpha_{\beta\gamma} = T^\alpha_{\beta\gamma} + S^\alpha_{\beta\gamma} + S^\alpha_{\gamma\beta},
\end{eqnarray}
is the cotorsion tensor, with $S^\alpha_{\beta\gamma} =
g^{\alpha\nu}g_{\beta\kappa}T^\kappa_{\nu\gamma}$, 
and $T^\alpha_{\beta\gamma} = \tilde\Gamma^\alpha_{\beta\gamma}
- \tilde\Gamma^\alpha_{\gamma\beta}$ the skew-symmetric torsion tensor.

Let us consider the Laplacian operator associated with this Cartan connection,
defined -in extending the usual definition- by
\begin{eqnarray}
H(\tilde\nabla) = 1/2{\rm trace}\tilde\nabla^2= 1/2g^{\alpha\beta}\tilde\nabla_\alpha \nabla_\beta;
\end{eqnarray}
 then, $\sigma (\tilde\nabla) = H(\tilde\nabla)$. A straightforward computation shows that 
that $H(\tilde\nabla)$ only depends in the trace of the torsion tensor and $g$:  
\begin{eqnarray}
H(\tilde\nabla) = 1/2\triangle_g + g^{\alpha\beta}Q_\beta\partial_\alpha \equiv H_0(g,Q),
\end{eqnarray}
with $Q = \tilde T^\nu_{\nu\beta}dx^\beta$, the trace-torsion one-form.

Therefore, for the Cartan connection $\tilde\nabla$ defined in $(10)$, we
have that
\begin{eqnarray}
\sigma (\tilde\nabla) = {1 \over 2}{\rm trace}(\tilde\nabla ^2) = {1 \over 2}\triangle _g +
\hat Q, 
\end{eqnarray}
with $\hat Q$ the vector-field conjugate to the $1$-form $Q$: $\hat Q (f) =
<Q, {\rm grad}~f>$, $f:M \to R$. In local coordinates,
\begin{eqnarray*}
\hat Q ^\alpha = g^{\alpha \beta}Q_\beta.
\end{eqnarray*}
We further have:
\begin{eqnarray}
X_0(\tilde\nabla) = B -{1 \over 2}g^{\alpha \beta}{\gamma\brace\alpha\beta}\partial_\gamma
-\hat Q,
\end{eqnarray}
Therefore, the invariant decomposition of $L$ is
\begin{eqnarray}
{1 \over 2}{\rm trace}(\tilde\nabla^2) +X_0(\tilde\nabla)  = {1 \over 2}\triangle _g + b,
\end{eqnarray}
with
\begin{eqnarray}
b = B - {1 \over 2}g^{\alpha\beta}{\gamma\brace\alpha\beta}\partial_\gamma.
\end{eqnarray}
Notice that $(15)$ can be thought as arising from a gauge transformation:
$\tilde b \to \tilde b - Q$, with $\tilde b$ the $1$-form conjugate to $b$.

If we take for a start $\tilde\nabla$ with Christoffel symbols of the form 
\begin{eqnarray}
\Gamma ^{\alpha}_{\beta \gamma} = {\alpha\brace\beta\gamma}~+~ {2 \over (n-1)}
\left\{\delta^\alpha_\beta ~Q_\gamma ~ - ~ g_{\beta\gamma}~Q^\alpha\right\}
\end{eqnarray} 
with
\begin{eqnarray}
Q = \tilde b, ~~~{\rm i.e.}~~~ \hat Q = b, 
\end{eqnarray}
we have in writing now $\nabla$ for the covariant derivative of $(18)$
\begin{eqnarray*}
X_0(\nabla) = 0,
\end{eqnarray*}
and
\begin{eqnarray}
H_0(g,Q) = \sigma (\nabla) = { 1\over 2}{\rm trace}(\nabla ^2) = {1 \over
2}{\rm trace}((\nabla ^g)^2) + \hat Q = {1 \over 2}\triangle _g + b. 
\end{eqnarray}
Therefore, for $\nabla$ as in $(18)$ we obtain a gauge theoretical invariant representation for $L$ given by
\begin{eqnarray}
L = H(\nabla)= {1\over 2} \nabla^2= \sigma (\nabla)  = {1 \over 2}{\rm trace}((\nabla^g)^2) + \hat Q  = H_0(g,Q). 
\end{eqnarray}

The restriction we have placed in the metric-compatible $\tilde\nabla$ to be as in $(18)$, i.e. only
the trace component of the irreducible decomposition of the torsion tensor
is taken, is due to the fact that all other components of this tensor do not
appear at all in the laplacian of (the otherwise too general) $\tilde\nabla$; in other words, 
$
H_0(g,Q) = {1 \over 2}(\nabla)^ 2 = {1 \over 2}(\tilde \nabla)^ 2 = {1 \over 2}\triangle_g + \hat Q,
$
with $\tilde\nabla$ given by $(10-11)$ and $\nabla$ given by $(18)$. In the particular
case of dimension $2$, this is automatically satisfied. In
the case we actually have assumed, $g$ is Riemannian, the
expression $(21)$ is the most general invariant laplacian (with zero potential term)
acting on
functions defined on a smooth manifold. 
This restriction, will allow us to establish a one-to-one correspondance between
Riemann-Cartan connections of the form $(18)$ with (generalized Brownian) diffusion processes.
These metric compatible connections we shall call RCW geometries (short for Riemann-Cartan-Weyl), since
the trace-torsion is a Weyl $1$-form [10]. Thus, these geometries do not have the historicity
problem which lead to Einstein's rejection of the first gauge theory ever proposed
by Weyl. We would like to remark that we first encountered these connections on developing a pre-symplectic structure for the
derivation of the dynamics of relativistic massive spinning systems subjected to exterior gravitational fields [57].

To obtain the most general form of the RCW laplacian, we only need to apply
to the trace-torsion one-form the
most general decomposition of one-forms on a smooth compact manifold. 
This amounts to give the constitutive equations of the particular
theory of fluctuations under consideration on the manifold $M$; see [22,26,49]. The answer
to this problem, is given by the well known de Rham-Kodaira-Hodge theorem, which
we present now.

We consider the Hilbert space of square summable 
$\omega$ of smooth differential forms of degree $k$ on $M$, with respect to ${\rm vol}_g$.
We shall denote this space as $L^{2,k}$. The
inner product is
\begin{eqnarray}
<<\omega ,\phi>>: = \int _M <\omega (x),\phi (x)>vol_g ,
\end{eqnarray}
where the integrand is given by the multiplication between the components 
$\omega_{\alpha_1 \ldots \alpha_k}$ of $\omega$
and the conjugate tensor:$g^{\alpha_1 \beta_1}\ldots g^{\alpha_k \beta_k}\phi_{\beta_1
\ldots \beta_k}$; alternatively, we can write in a coordinate independent way: $<\omega (x),\phi(x)>vol_g = \omega (x) \wedge *\phi(x)$, 
with $*$ the Hodge star operator, for any $\omega ,\phi \in L^{2,k}$.

The de Rham-Kodaira-Hodge operator on $L^{2,k}$ is defined as 
\begin{eqnarray}
\triangle_k =  - (d + \delta)^2 = -(d \delta + \delta d), 
\end{eqnarray}
where $\delta$ is the formal adjoint defined on $L^{2,k+1}$ of the exterior differential
operator $d$ defined on $L^{2,k}$:
\begin{eqnarray*}
<<\delta \phi ,\omega>> = <<\phi ,d\omega>>,
\end{eqnarray*}
for $\phi \in L^{2,k+1}$ and $\omega \in L^{2,k}$. Then, $\delta^2 = 0$.

Let $R:(TM \oplus TM)\oplus TM \to TM$ be 
the (metric) curvature tensor defined by: $
(\nabla^g)^2 Y (v_1,v_2) = (\nabla^g)^2 Y(v_2,v_1) + R(v_1,v_2)Y(x)$.
From the Weitzenbock formula [14] we have
\begin{eqnarray*}
\triangle _1 \phi (v) = {\rm trace}~(\nabla^g)^2 \phi (-,-) - Ric_x (v,\hat\phi_x),
\end{eqnarray*}
for $v \in T_xM$ and $Ric_x (v_1,v_2) = {\rm trace}~<R(-,v_1)v_2,->_x.$
Then,  $\triangle_0 = (\nabla^g)^2 = \triangle_g$ so that in the case of $k = 0$,
the de Rham-Kodaira operator coincides with the Laplace-Beltrami operator on functions. 

The de Rham-Kodaira-Hodge theorem states that $L^{2,1}$ admits the following invariant
decomposition. Let $\omega \in L^{2,1}$; then,
\begin{eqnarray}
\omega =  d~f + A_1 + A_2, 
\end{eqnarray}
where $f:M \to R$ is a smooth function on $M$, $A_1$ is a co-closed smooth $1$-form:
$\delta A_1 = -div_g \hat A_1 = 0$,
and $A_2$ is a co-closed and closed smooth $1$-form:
\begin{eqnarray}
\delta A_2 = 0, dA_2 = 0. 
\end{eqnarray}
Otherwise stated, $A_2$ is an harmonic one-form, i.e. 
\begin{eqnarray}
\triangle_1 A_2 = 0. 
\end{eqnarray}
Furthermore, this decomposition is orthogonal in $L^{2,1}$, i.e.:
\begin{eqnarray}
<<df,A_1>> = << df,A_2>> = <<A_1,A_2>> = 0. 
\end{eqnarray}
 
{\bf Remark~1.} Note that $A_1 + A_2$ is itself a co-closed one-form. If
we consider an augmented configuration space $R \times M$ for an incompressible
fluid, this last decomposition will be the fluid's velocity. If we 
consider instead a four-dimensional Lorentzian manifold provided with a Dirac-Hestenes
spinor operator field (DHSOF), one needs the whole decomposition $(24)$ associated
to an invariant density $\rho$ of the diffusion (i.e. a solution of the equation $H_0(g,Q)^\dagger\rho = 0$)  to describe two electromagnetic
potentials such that when restricted to the spin-plane of the DHSOF, they enforce the
equivalence between the Dirac-Hestenes equation for the DHSOF on a manifold
provided with a RCW connection, and the free Maxwell equation on the Lorentzian
manifold; see [26, 64]. 

\section{Generalized Laplacians on Differential Forms}

\hspace{2em} Consider the family of zero-th order differential operators acting on smooth $k$-forms, i.e.
differential forms of degree $k$ ($k = 0,\ldots,n$) defined on $M$:
\begin{eqnarray}
H_k(g,Q): = 1/2\triangle_k + L_{\hat Q}, 
\end{eqnarray}
The second term in $(28)$
denotes the Lie-derivative with respect to the vectorfield $\hat Q$. Recall that the
Lie-derivative is independant of the metric:for any smooth vectorfield $X$ on $M$
\begin{eqnarray}
L_X = i_Xd + di_X,, 
\end{eqnarray}
where $i_X$ is the interior product with respect to $X$: for arbitrary 
vectorfields $X_1,\ldots,X_{k-1}$ and $\phi$ a $k$-form defined on $M$, we have $(i_X \phi)(X_1,\ldots,X_{k-1})
= \phi (X,X_1,\ldots,X_{k-1})$. Then, for $f$ a scalar field, $i_X f = 0$
and
\begin{eqnarray}
L_{X}f = (i_{X}d + di_{X})f = i_{X} df = g(\tilde X,df) = X(f).
\end{eqnarray}
where $\tilde X$ denotes the $1$-form associated to a vectorfield $X$ on $M$ via
$g$. We shall need later the following identities between operators acting on smooth $k$-forms,
which follow easily from algebraic manipulation of the definitions: 
\begin{eqnarray}
d\triangle_k = \triangle_{k+1}d,~k=0,\ldots,n, 
\end{eqnarray}
and
\begin{eqnarray}
\delta \triangle_ k = \triangle_{k-1}\delta ,~ k=1,\ldots,n,
\end{eqnarray}
and finally, for any vectorfield $X$ on $M$ we have that $dL_X = L_Xd$ and therefore 
\begin{eqnarray}
dH_k(g,Q) = H_{k+1}(g,Q)d,~ k=0,\ldots,n.
\end{eqnarray}

In $(28)$ we retrieve for scalar fields $(k=0)$ the operator $H(g,Q)$ defined in $(21)$.

{\bf Proposition~1:} Assume that $g$ is non-degenerate. There is a one-to-one mapping 
$$
\nabla \leadsto H_k(g,Q) = 1/2\triangle_k + L_{\hat Q}
$$
between the space of $g$-compatible affine connections $\nabla$ with Christoffel coefficients of the form
\begin{eqnarray}
\Gamma ^{\alpha}_{\beta \gamma} = {\alpha\brace\beta\gamma}~+~ {2 \over (n-1)}
\left\{\delta^\alpha_\beta ~Q_\gamma ~ - ~ g_{\beta\gamma}~Q^\alpha\right\}
\end{eqnarray} 
and the space of elliptic second order differential operators on $k$-forms ($k=0,\ldots,n)$
with zero potential term.

\section{Riemann-Cartan-Weyl Connections and the Laplacians for Differential Forms}
\hspace{2em} In this section we shall construct the diffusion processes for scalar fields.

In the following we shall further assume that $Q = Q(\tau,x)$ is a time-dependant $1$-form, so that
we have a time-dependant RCW connection on $M$, which we think of as a space manifold.
The stochastic flow associated to the
diffusion generated by $H_0(g,Q)$ 
has for sample paths the continuous
curves $\tau \mapsto x_{\tau}\in M$ satisfying the Ito invariant non-degenerate
s.d.e. (stochastic differential equation)
\begin{eqnarray}
dx(\tau) = X(x(\tau))dW(\tau) + \hat Q(\tau,x(\tau))d\tau.
\end{eqnarray}
In this expression, the diffusion tensor $X = (X^\alpha _\beta (x))$ is a linear surjection
$X(x):R^m \rightarrow T_xM$ satisfying $X^\alpha_\nu X^\beta _\nu = g^{\alpha \beta}$,
and $\{W(\tau),\tau \ge 0\}$ is a standard Wiener process on $R^n$. Thus $<W_\tau> = 0$
and $<W^i_\tau W^j_\tau> = \delta_{ij}\tau$, where $<->$ denotes expectation with
respect to the zero-mean standard Gaussian function on $R^m$ ($m \ge n$).
Here $\tau$ denotes the time-evolution parameter of the diffusion (in a relativistic
setting it should not be confused with the time variable), and for simplicity we shall assume
always that $\tau \ge 0$.
Consider the canonical Wiener space $\Omega$ of continuous
maps $\omega :R \to R^n, \omega (0) = 0$,
with the canonical realization of the Wiener process $W(\tau)(\omega) = \omega (\tau)$. 
The (stochastic) flow of the s.d.e. $(35)$ is a 
mapping 
\begin{eqnarray}
F_\tau :M\times \Omega\to M, ~~\tau \ge 0,
\end{eqnarray}
such that for each $\omega\in\Omega$,
the mapping $F_.(.~,\omega):[0,\infty)\times M\to M,$
is continuous and such that $\{F_\tau(x):\tau\ge 0\}$
is a solution of equation $(35)$ with $F_0(x)=x$, for any $x\in M$.

Let us assume in the following that the components $X^\alpha _\beta$, $\hat Q^\alpha$, $\alpha,\beta=1,\ldots ,n$
of the vectorfields $X$ and $\hat Q$ on $M$ in $(35)$ are predictable functions which further belong
to $C^{m,\epsilon}_b$ ($0\le \epsilon \le 1$, $m$ a non-negative integer),
the space of Hoelder bounded continuous functions of degree $m \ge 1$ and exponent $\epsilon$, and 
also that
$\hat Q^\alpha (\tau) \in L^1(R)$, for any
$\alpha =1,\ldots,n$. With these regularity conditions, if we further assume that $\{x(\tau): \tau \ge 0\}$
is a semimartingale (see Appendix) on a probability space $(\Omega,{\cal F},P)$, then it follows from Kunita
[24] that $(35)$ has a modification (which with abuse of notation we denote as)
\begin{eqnarray}
F_\tau(\omega):M\to M,~~~ F_\tau(\omega)(x)=F_\tau(x,\omega),
\end{eqnarray}
which is a diffeomorphism  of class $C^m$, almost surely for  $\tau\ge  0$ and $\omega\in\Omega$.  
We can obtain an identical result if we assume instead Sobolev regularity conditions. Indeed, 
assume that the components of $\sigma$ and $\hat Q$, $\sigma_i^\beta \in H^{s+ 2}(T^*M)$ and $\hat Q^\beta \in H^{s+1}(T^*M)$, $1 \le i \le m$, $1 \le \beta
\le n$, where the Sobolev
space $H^s(T^*M) = W^{2,s}(T^*M)$ with $s > {n \over 2}+m$ [52]. Then, the flow of $(35)$ for fixed $\omega$ defines a diffeomorphism
in $H^s(M,M)$ (see [53]), and hence by the Sobolev embedding theorem, a diffeomorphism in $C^m(M,M)$ (i.e. a mapping from $M$ to $M$
which is $m$-times continuously differentiable as well as its inverse.)   In
any case, for $1 \le m$ we can consider the Jacobian (¨velocity¨) flow of $\{x_\tau: \tau \ge 0\}$. It is a
random diffusion process on $TM$, the tangent bundle of $M$.

{\bf Remarks~2:} In the differential geometric approach -pioneered by V. Arnold- for integrating NS
on a smooth manifold as a perturbation (due to the diffusion term we shall present below)
of the geodesic flow in
the group of volume preserving diffeomorphisms of $M$ (as the solution of the Euler
equation), it was proved that under the above regularity conditions on the initial velocity, the solution
flow of NS defines a diffeomorphism in $M$ of class $C^m$; see Ebin and Marsden
[9]. The difference of this classical approach with the one presented here, is to integrate NS through a time-dependant
${\bf random}$ diffeomorphism associated with a RCW connection. As wellknown, these regularity conditions are basic in 
the usual functional analytical treatment of NS pioneered by Leray [45] (see also Temam [7]), and they are further related to the multifractal
structure of turbulence [41].
This diffeomorphism property of random flows is fundamental for the construction 
of their ergodic theory (provided an invariant measure for the processes exists), and in particular, of 
quantum mechanics amd non-linear non-equilibrium thermodynamics [10,21,22,26,49].

Let us describe now the Jacobian flow.
We can describe it as the stochastic
process on the tangent bundle, $TM$, given by $\{v(\tau) := T_{x_0}F_\tau(v(0)) \in T_{F_\tau(x_0)}M,\break v(0) \in T_{x_0}M\}$; here $T_zM$
denotes the tangent space to $M$ at $z$ and $T_{x_0}F_\tau$ is the linear derivative
of $F_\tau $ at $x_0$. The process $\{v_\tau, \tau \ge 0\}$ can be described (see [27]) as the solution of the invariant Ito
s.d.e. on $TM$:
\begin{eqnarray}
dv(\tau) = \nabla^ g\hat Q(\tau,v(\tau))d\tau + \nabla^gX(v(\tau))dW(\tau)
\end{eqnarray}
If we take $U$ to be an open
neighborhood in $R^n$ so that $TU = U\times R^n$, then $v(\tau) = (x(\tau),\tilde v(\tau))$
is described by the system given by integrating $(35)$ and the covariant Ito s.d.e.
\begin{eqnarray}
d\tilde v(\tau)(x(\tau)) = \nabla^g X(x(\tau))(\tilde v(\tau))dW(\tau) + \nabla^g \hat Q(\tau,x(\tau))(\tilde v(\tau))
d\tau,
\end{eqnarray}
with initial condition $\tilde v(0)=v_0 \in T_{x(0)}$.
Thus, $\{v(\tau) =(x(\tau),\tilde v(\tau)), \tau\ge 0\}$ defines a random flow on $TM$.

${\bf Theorem ~2:}$ For any differential $1$-form
$\phi$ of class $C^{1,2}(R\times M)$ (i.e. in a local coordinate system $\phi = a_\alpha (\tau) dx^\alpha$,
with $a_\alpha (\tau,.)\in C^2(M)$ and $a_\alpha (.,x) \in C^1(R)$) we have the Ito formula (Corollary 3E1 in [27]):

\begin{eqnarray}
\phi (v_\tau)& = &\phi(v_0) + \int^\tau_0 \phi (\nabla^g X(v_s)dW_s
+ \int _0 ^\tau [{\partial \over \partial s}
+ H_1(g,Q)]\phi(v_s)ds\nonumber\\& + & \int^\tau_0 \nabla^g \phi (X(x)dW_s)(v_s)\nonumber\\
& + &\int^\tau_0 {\rm trace}~d\phi (X(x_s)-,\nabla ^g X(v_s))(-)ds  
\end{eqnarray}

In the last term in $(40)$ the trace is taken in the argument $-$ of the bilinear
form and further we have the mappings
$$
\nabla^g Y: TM \to TM;
\nabla^g \phi:TM \to T^*M.
$$

${\bf Remarks~3:}$ From $(40)$ we conclude that the infinitesimal generators (i.g., for short in the following)  of the
derived stochastic process is not
$\partial_\tau + H_1(g,Q)$, due to the last term in $(40)$. 
This term vanishes identically in the case we shall present in the following
section, that of gradient diffusions. An alternative method which bypasses the velocity
process is the construction of the
generalized Hessian flow further below. Both methods will provide for the setting for the integration of the Navier-Stokes
equations.

\section{Riemann-Cartan-Weyl Gradient Diffusions}
\hspace{2em}Suppose that there is an isometric embedding of an $n$-dimensional 
compact orientable manifold $M$ into a Euclidean space
$R^m$:$f: M \to R^m, f(x) = (f^1(x),\ldots,f^m(x))$. Suppose further that $X(x): R^m \to T_xM$, 
is the orthogonal projection  of $R^m$ onto $T_xM$ the tangent space at $x$ to $M$, considered as
a subset of $R^m$. Then, if $e_1,\ldots,e_m$ denotes the standard basis of $R^m$, we have 
\begin{eqnarray}
X = X^ie_i, ~{\rm with}~ X^i = {\rm grad}~f^i, i =1,\ldots,m.
\end{eqnarray}
The second fundamental form [25] is a bilinear symmetric map
\begin{eqnarray}
\alpha_x :T_xM \times T_xM \to \nu_xM, x\in M,
\end{eqnarray}
with $\nu_x M = (T_xM)^\perp$ the space of normal vectors at $x$ to $M$. We then have the
associated mapping
\begin{eqnarray}
A_x: T_xM \times \nu_xM \to T_xM, <A_x (u,\zeta),v>_{R^m} = <\alpha_x(u,v),\zeta>_{R^m},
\end{eqnarray}
for all $\zeta \in \nu_xM$, $u,v \in T_xM$. Let $Y(x)$ be the orthogonal projection
onto $\nu_xM$
\begin{eqnarray}
Y(x) = e - X(x)(e), x \in M, e \in R^m.
\end{eqnarray}
Then: 
\begin{eqnarray}
\nabla^g X(v)(e) = A_x(v,Y(x)e), v\in T_x M, x \in M.
\end{eqnarray}
For any $x \in M$, if we take $e_1,\ldots,e_m$ to be an orthonormal base for $R^m$ such that
$e_1,\ldots,e_m \in T_xM$, then for any $v \in T_xM$ ,we have 
\begin{eqnarray}
{\rm either}~\nabla^g X(v)e_i = 0, ~{\rm or}~X(x)e_i = 0.
\end{eqnarray}

We shall consider next the RCW gradient diffusion processes, i.e. for which in equation $(35)$
we have specialized taking $X= {\rm grad}f$. Let $\{v_\tau:\tau \ge 0 \}$ be the associated derived ¨velocity¨
process.
We shall now give the Ito-Elworthy formula for $1$-forms.

${\bf Theorem~3.}$ Let $f:M \to R^m$ be an isometric embedding. For any differential form $\phi$ of degree $1$ in $C^{1,2}(R\times M)$, the Ito formula is
\begin{eqnarray}
\phi(v_\tau) & = & \phi (v_0) + \int^\tau_0 \nabla ^g \phi(X(x_s)dW_s)v_s + \int^\tau_0
\phi (A_x(v_s, Y(x_s)dW_s)\nonumber\\
& + & \int^\tau_0 [{\partial \over \partial s}+ H_1(g,Q)]\phi(v_s)ds,
\end{eqnarray}
i.e. $\partial_\tau + H_1(g,Q)$, is the i.g. (with domain the differential $1$-forms belonging to
$C^{1,2}(R\times M)$) of $\{v_\tau : \tau \ge 0 \}$.

Proof: It follows immediately from the facts that the last term in the r.h.s. of
$(40)$ vanishes due to $(46)$, while the second term in the r.h.s. of
$(40)$ coincides with the third term in $(47)$ due to $(45)$. 

Consider
the value $\Phi_x$ of a $k$-form at $x \in M$ as a linear map: $\Phi_x: \Lambda^k T_xM \to R$.
In general, if $E$ is a vector space and $A:E \to E$ is a linear map, we have the 
induced maps
\begin{eqnarray*}
\Lambda^k A: \Lambda^k E \to \Lambda^k E,~
\Lambda^k (v^1\wedge \ldots \wedge v^k): = Av^1\wedge \ldots \wedge Av^k;
\end{eqnarray*}
and
\begin{eqnarray*}
(d\Lambda^k)A :\Lambda^k E \to \Lambda^k E,~
\end{eqnarray*}
\begin{eqnarray*}
(d\Lambda^k)A(v^1\wedge \ldots \wedge v^k): = \sum_{j=1}^k &
v^1\wedge \ldots \wedge v^{j-1}\wedge Av^j \wedge v^{j+1} \wedge \ldots \wedge v^k.
\end{eqnarray*}
For $k= 1, (d\Lambda)A = \Lambda A$. The Ito formula for $k$-forms, $1 \le k \le n$, is due to Elworthy (Prop. 4B
[27]). 

${\bf Theorem~4.}$ Let $M$ be isometrically embedded in $R^m$. Let $V_0 \in \Lambda^k T_{x_0}M$. Set $V_\tau = \Lambda ^k (TF_\tau)(V_0)$
Then for any differential form $\phi$ of degree $k$ in $C^{1,2}(R\times M)$, $ 1 \le k \le n$,
\begin{eqnarray}
\phi (V_\tau)&=& \phi (V_0) + \int^\tau _0 \nabla^g \phi (X(x_s)dWs)(V_s) \nonumber\\
& + &\int ^\tau_0 \phi((d\Lambda)^k A_{x_s}(-,Y(x_s)dW_s)(V_s))\nonumber\\
& + &\int ^\tau_0 [{\partial \over \partial
s}+H_k(g,\hat Q)]\phi(V_s)ds
\end{eqnarray}
i.e., 
${\partial_\tau}+ H_k(g,\hat Q)$ is the i.g. (with domain of definition
the differential forms of degree $k$ in $C^{1,2}(R\times M)$) of $\{V_\tau :\tau \ge 0\}$.

{\bf Remarks 4:} Therefore, starting from the flow $\{F_\tau: \tau \ge 0\}$ of the s.d.e. $(35)$ with i.g. given by
$\partial _\tau + H_0(g,Q)$ , we obtained that the derived velocity process $\{v(\tau): \tau \ge 0 \}$
given by $(38)$ (or $(35)$ and $(39)$) has $H_1(g,Q)$ as i.g.; finally, if we consider the diffusion
of differential forms of degree $k \ge 1$, we get that $\partial_\tau + H_k(g,Q)$ is the i.g. 
of the process $\Lambda^k v(\tau)$, i.e. the exterior product of degree $k$ ($k=1,\ldots,n$)
of the velocity process.
In particular, $\partial_\tau + H_2(g,Q)$ is the i.g. of the stochastic process
$\{v(\tau)\wedge v(\tau): \tau \ge 0\}$.

Note that consistently with the notation we have that $\{\Lambda^0 v_\tau: \tau \ge 0\}$ is the 
position process $\{x_\tau: \tau \ge 0\}$ untop of which $\{\Lambda^k v_\tau:\tau \ge 0\}$, ($1 \leq k \leq n$)
is fibered (recall, $\Lambda^0 (M) = M$).
We can resume our results in the following theorem.

{\bf Theorem 5.} Assume $M$ is isometrically embedded in $R^m$. There is a one to
one correspondance between RCW connections $\nabla$ determined by a Riemannian metric
$g$ and trace-torsion $Q$ with the family of gradient diffusion processes $\{\Lambda^k v_\tau: \tau \ge 0\}$
generated by $H_k(g,Q)$, $k=0,\ldots,n$

Finally, we are now in a situation for presenting the solution of the Cauchy problem
\begin{eqnarray}
{\partial \phi \over \partial \tau} = H_k(g,Q_\tau)(x)\phi, \tau \in [0,T]
\end{eqnarray}
with the given initial condition
\begin{eqnarray}
\phi(0,x) = \phi_0(x),
\end{eqnarray}
for $\phi$ and $\phi_0$ $k$-forms on a smooth compact orientable manifold isometrically embedded in $R^m$.
From the Ito-Elworthy formula follows that the formal solution of this problem is as follows: Consider the diffusion process
on $M$ generated by $H_0(g,Q)$: For each $\tau \in [0,T]$ consider the s.d.e. (with $s \in [0,\tau]$):
\begin{eqnarray}
dx^{\tau ,x}_s = X(x^{\tau ,x}_s)dW_s +\hat Q(\tau -s,x^{\tau ,x}_s)ds,
\end{eqnarray} 
with initial condition
\begin{eqnarray}
x^{\tau ,x}_0 = x,
\end{eqnarray}
and the derived velocity process $\{v^{\tau ,v(x)}_s = x^{\tau ,x}_s,\tilde v^{\tau ,v(x)}_s, 0 \le s \le \tau\}$:
\begin{eqnarray}
d\tilde v^{\tau ,v(x)}_s = \nabla^g X(x^{\tau ,x}_s)
(\tilde v^{\tau ,v(x)}_s)dW_s +\nabla^g\hat Q(\tau -s, x^{\tau ,x}_s)(\tilde v^{\tau ,v(x)}_s)ds, 
\end{eqnarray}
with initial condition
\begin{eqnarray}
\tilde v^{\tau ,v(x)}_0 = v(x).
\end{eqnarray}
Then, the $C^{1,2}$ (formal) solution of the Cauchy problem defined in $[0,T]\times M$ is
\begin{eqnarray}
\phi(\tau,x)(\Lambda^k v(x)) = E_x[\phi_0(x^{\tau,x}_\tau)(\Lambda^k \tilde v^{\tau,v(x)}_\tau].
\end{eqnarray}.

\section{The Navier-Stokes Equation and Riemann-Cartan-Weyl Diffusions}
\hspace{2em}In the sequel, $M$ is a compact orientable (possible with smooth boundary
$\partial M$) $n$-manifold with a Riemannian metric $g$.
We provide $M$ with a $1$-form whose de Rham-Kodaira-Hodge (RKH for short) decomposition is
\begin{eqnarray*}
Q(x) = df(x) + u(x),~~\delta u = -div (\hat u) = 0,
\end{eqnarray*}
where $f$ is a scalar field and $u$ is a coclosed $1$-form, weakly orthogonal to 
$df$, i.e. $\int g(df,u) vol_g = 0$. We shall assume that $u(x,0) = u(x)$ is the initial velocity $1$-form of an
incompressible viscous fluid on $M$, and that we further have a $1$-form $Q(x,\tau) = Q_\alpha (x,\tau)dx^\alpha$ whose RKH decomposition is:
\begin{eqnarray}
Q(x,\tau)= df(x,\tau) + u(x,\tau),  
\end{eqnarray}
with $\delta u_\tau (x) =\delta u(x,\tau) =0$ (incompressibility condition), and
\begin{eqnarray*}
\int g(df_\tau,u_\tau)vol(g) = 0,
\end{eqnarray*}
which satisfies the evolution equation on $M \times R$ (Eulerian representation of the fluid):
\begin{eqnarray}
{\partial Q _\alpha \over \partial \tau} + \nabla ^g_{\hat u}Q_\alpha  = -Q_\beta \nabla^g_\alpha u^\beta +
\nu \triangle_1 Q_\alpha,
\end{eqnarray}
Here $\nu$ is the kinematical viscosity. In the above notations and in the
following, all covariant
operators act in the $M$ variables only. In the formulation of Fluid Mechanics in Euclidean domains,
$Q(x,\tau)$ receives the name of (Buttke) "magnetization variable" [1]. 

{\bf Remarks~5:} We recall that to take the RKH decomposition of the velocity
of a viscous fluid is a basic procedure in Fluid Mechanics [1,6,7,9]. 
We shall see below that $Q$ and in particular $u$ are related to a natural
RCW geometry of the incompressible fluid.
In the formulation of Quantum Mechanics and
of non-linear non-equilibrium thermodynamics stemming from RCW diffusions, we have a RKH
decomposition of the trace-torsion associated to a stationary state; see [10,21,22,26,49]. 
This decomposition allows to associate with the divergenceless
term of the trace-torsion a probability current which
characterizes the time-invariance symmetry breaking of the diffusion process, and is central to the construction of the ergodic theory
of these flows.

Equation $(57)$ is the gauge-invariant form of the
NS for the velocity $1$-form $u(x,\tau)$.
Indeed, if we substitute
the decomposition of $Q(x,\tau) = Q_\tau(x)$ into $(57)$
we obtain,
\begin{eqnarray}
{\partial u \over \partial \tau} + \nabla^g_{\hat u_\tau}u_\tau = \nu \triangle_1 u_\tau - 
d({\partial f \over \partial \tau} +
\nabla^g _{\hat u_\tau}f + {1 \over 2}|u_\tau|^2 - \nu \triangle_g f).
\end{eqnarray}
Consider the operator $P$ of projection of $1$-forms into co-closed $1$-forms:
$P\omega = \alpha$ for any one-form $\omega$ whose RKH decomposition is
$\omega = df + \alpha, ~{\rm with}~ \delta \alpha =0$.
From $(32)$ we get that
\begin{eqnarray}
P\triangle_1 u_\tau = \triangle_1 u_\tau, 
\end{eqnarray}
and further applying $P$ to $(57)$ we finally get the well known covariant NS (with no exterior forces; the gradient
of the pressure
term disappears by projecting with $P$
[1,9])
\begin{eqnarray}
{\partial u \over \partial \tau} + P[\nabla^g_{\hat u_\tau}u_\tau] - \nu\triangle_1 u_\tau = 0.
\end{eqnarray}
Conversely, starting with equation $(58)$ which is equivalent to NS we obtain $(57)$. Note
that $Q_\tau$ and $u_\tau$ differ by a differential of a function for all times. Multiplication
of $(58)$ by $I-P$ (I the identity operator) yields an equation for the evolution of $f$ which is only arbitrary
for $\tau = 0$.
Now we note that the non-linearity of NS originates from applying $P$ to the term
\begin{eqnarray*}
\nabla^g_{\hat u_\tau}u_\tau = i_{\hat u_\tau}du_\tau, 
\end{eqnarray*}
which taking in account $(29)$ can still be written as
\begin{eqnarray}
L_{\hat u_\tau}u_\tau - di_{\hat u_\tau}u_\tau = L_{\hat u_\tau}u_\tau - 1/2d(|u_\tau|^2).
\end{eqnarray}
Applying $P$ to $(61)$, we see that the kinetic energy term there disappears
and the non-linear term in NS can be written as
\begin{eqnarray}
P[\nabla^g_{\hat u_\tau}u_\tau] = P[\L_{\hat u_\tau}u_\tau].
\end{eqnarray}
Therefore, from $(28)$ and $(62)$, NS $(60)$ takes the final concise form
\begin{eqnarray}
{\partial u \over \partial \tau} = PH_1(2\nu g,{-1 \over 2\nu} u_\tau)u_\tau.
\end{eqnarray}
Therefore we have found that NS for the velocity of an incompressible fluid is a 
a non-linear diffusion equation.  

Let us introduce the vorticity two-form
\begin{eqnarray}
\Omega_\tau = du_\tau.
\end{eqnarray}
Note that also $\Omega_\tau = dQ_\tau$. Now, if we know $\Omega_\tau$ for any $\tau \ge 0$, we can obtain
$u_\tau$ (or still $Q_\tau$) by inverting the definition $(64)$. Namely, applying $\delta$
to $(64)$ and taking in account $(23)$ we obtain
the Poisson-de Rham equation (would $g$ be hyperbolic, it is the Maxwell-de Rham equation
[10a])
\begin{eqnarray}
\triangle_1 u_\tau = -\delta \Omega_\tau.
\end{eqnarray}
and an identical equation for $Q_\tau$. (Note that if we know $Q_\tau$ we can reconstruct
$f_\tau$ by solving $-\delta Q_\tau = div (\hat Q_\tau) = \triangle_g f_\tau$, for any $\tau$.) From the Weitzenbock
formula, we can write $(65)$ showing the coupling 
of the Ricci metric curvature to the velocity $u = u_\alpha (x,\tau)dx^\alpha$:
\begin{eqnarray}
(\nabla^g)^2 u_\tau - R_{\alpha \beta}u_\tau^\beta dx^\alpha = -\delta \Omega_\tau.
\end{eqnarray}
with $R_\alpha ^\beta (g) = R_{\mu \alpha}^{~~~\mu \beta}(g)$, the Ricci (metric) 
curvature tensor. Thus, the vorticity $\Omega_\tau$ is a source for the velocity one-form
$u_\tau$, for all $\tau$; in the case that $M$ is a compact euclidean domain, equation
$(66)$ is integrated to give the Biot-Savart law of Fluid Mechanics [1,39].

Now, apply $d$ to $(63)$ and further RKH decompose $L_{-\hat u_\tau}u_\tau = \alpha_\tau + dp_\tau$
(with $p_\tau$ the pressure at time $\tau$); in account
that 
\begin{eqnarray*}
dPL_{-\hat u_\tau}u_\tau
= d\alpha_\tau = d(\alpha_\tau + dp_\tau) = dL_{-\hat u_\tau}u_\tau = L_{-\hat u_\tau}du_\tau = L_{-\hat u_\tau}\Omega_\tau,
\end{eqnarray*}
and that from $(31)$ we have that $d\triangle_1u_\tau = \triangle_2\Omega_\tau$, we therefore obtain the linear evolution equation
\begin{eqnarray}
{\partial \Omega_\tau \over \partial \tau} = H_2(2\nu g,{-1 \over 2\nu}u_\tau)\Omega_\tau.
\end{eqnarray}

Thus, we have proved that the Navier-Stokes equation is a linear diffusion equation generated by a RCW connection. 
This connection has $2 \nu g$ for the
metric, and the time-dependant trace-torsion of this connection is $-u/2\nu$.
Then,
the drift of this process does not depend explicitly on $\nu$, 
as it coincides with the vectorfield associated via $g$ to $-u_\tau$, i.e.$-\hat u_\tau$. Notice that when $\nu$ tends to zero, i.e. the Euler equation, the trace-torsion becomes singular.

${\bf Theorem~6:}$ Given a compact orientable Riemannian
manifold with metric $g$, the Navier-Stokes equation
$(63)$ for an incompressible fluid
with velocity one-form $u=u(\tau,x)$ such that $\delta u_\tau =0$, assuming sufficiently regular conditions,
are equivalent to a linear diffusion process for the vorticity given by $(67)$ with $u_\tau$ satisfying
the Poisson-de Rham equation $(65)$. The RCW connection on $M$
generating this process is determined by the metric $2\nu g$ and a trace-torsion $1$-form
given by $-u/2\nu$.

{\bf Remarks 6:} We would like to recall that in the gauge theory of gravitation [46,57] the torsion is
related to the translational degrees of freedom present in the Poincare group, i.e. to the gauging of
momentum. Here we find a similar, yet dynamical situation, in which the trace-torsion is related to the
velocity. We would like to point out further, than on setting the skew-symmetric torsion to be zero
on taking RCW connnections, due to the fact that only the trace-torsion appears in the laplacian
generating generalized Brownian
motions, we are setting to zero the inertial fields which can be associated with the skew-symmetric torsion.
Thus, generalized Brownian motions are not  generated by inertial fields [64].

\section{\bf Random Diffeomorphims and the Navier-Stokes Equations}
\hspace{2em}In the following we assume additional conditions on $M$, namely that $f:M \rightarrow R^m$
is an isometric embedding, and that $M$ has no boundary. 

Let $u$ denote a solution of $(63)$ (which we assume that exists for $\tau \in [0,T]$),
and consider the flow $\{F^\nu _\tau: \tau \ge 0\})$  
of the s.d.e. whose i.g. is ${\partial \over \partial \tau}+ H_0(2\nu g,{-1 \over 2\nu}u)$; from $(35)$ we
know that this is the flow defined by integrating the non-autonomous Ito s.d.e. with $X = \nabla f$)
\begin{eqnarray}
dx^{\nu,\tau,x} = [2\nu]^{1 \over 2}X(x^{\nu,\tau,x})dW(\tau)
 -\hat u(\tau,x^{\nu,\tau,x})d\tau, x^{\nu,0,x}=x, \tau \in [0,T]. 
\end{eqnarray}

We shall assume in the following that the diffusion tensor $X$ and the drift $\hat u_\tau$ have the regularity
conditions stated in Section $4$, so that the randoms flows of $(68)$ is a diffeomorphism of $M$ of class $C^m$, $m \ge 1$. 

{\bf Theorem~7:} Equation $(68)$ is a random Lagrangian representation for the fluid particles positions, i.e. $x(\tau)$ is the random
position of the fluid particles of the incompressible fluid whose velocity obeys $(63)$.

{\bf Remark~7:} Note that the drift of the derived process $\{\tilde v_\tau: \tau \ge 0\}$ is minus the
deformation tensor of the fluid (see equation $(71)$ further below).  This will have a crucial role in the solution for the vorticity equation as well
as the kinematic dynamo problem. We further note that if in $(68)$ we set the viscosity to zero, we get the
classical flow of the Euler equation.

\subsection{\bf Cauchy Problem for the Vorticity} 
\hspace{2em} Let us solve the Cauchy problem for $\Omega (\tau,x)$ of class $C^m$ in $[0,T]\times M$ satisfying
$(67)$ with initial condition $\Omega_0(x)$.

For each $\tau \in [0,T]$ consider the s.d.e. (with $s \in [0,\tau]$) (obtained by running backwards the
Lagrangian representation $(68)$ above):
\begin{eqnarray}
dx^{\nu,\tau ,x}_s = (2\nu)^{1 \over 2}X(x^{\nu,\tau ,x}_s)dW_s -\hat u(\tau -s,x^{\nu,\tau ,x}_s)ds,
\end{eqnarray} 
with initial condition
\begin{eqnarray}
x^{\nu,\tau ,x}_0 = x,
\end{eqnarray}
and the derived velocity process $\{v^{\nu ,\tau ,v(x)}_s = (x^{\nu,\tau ,x}_s,\tilde v^{\nu,\tau ,v(x)}_s), 0 \le s \le \tau\}$:
\begin{eqnarray}
d\tilde v^{\nu,\tau ,v(x)}_s = (2\nu)^{1 \over 2}\nabla^g X(x^{\nu,\tau ,x}_s)
(\tilde v^{\nu ,\tau ,v(x)}_s)dW_s -\nabla^g\hat u(\tau -s, x^{\nu,\tau ,x}_s)(\tilde v^{\nu,\tau ,v(x)}_s)ds, 
\end{eqnarray}
with initial condition
\begin{eqnarray}
\tilde v^{\nu,\tau ,v(x)}_0 = v(x).
\end{eqnarray}

{\bf Theorem 8:}
Let $\Omega_\tau(x)$ be a bounded $C^{1,2}$ solution of the Cauchy problem; then it follows from the Ito formula-Elworthy $(48)$
(with $k=2$) is 
\begin{eqnarray}
\tilde \Omega_\tau (\Lambda^2 v(x))= E_x[\Omega_0(x^{\nu,\tau ,x}_\tau)(\Lambda^2\tilde v^{\nu ,\tau ,v(x)}_\tau)]
\end{eqnarray}
where the expectation value at $x$ is taken with respect to the measure on the
process $\{x^{\nu,\tau,x}_\tau: \tau \in [0,T]\}$ (whenever it exists):

{\bf Proof:} It follows just from applying the Ito-Elworthy formula for $2$-forms.c.q.d.

{\bf Remarks 8:} We would like to examine the physical interpretation of the representation $(73)$. To determine the vorticity at time
$\tau$ on a point $x$ evaluated on a bivector $\Lambda^ 2 v(x)$, we run backwards the random lagrangian representations starting at time $\tau$ at $x$, and its Jacobian flow starting at $v(x)$ along which we transport the time-zero vorticity, and further we take the mean value along all possible random paths. Furthermore, the transport of $\Omega_0$ along this Jacobian flow, indicates that the original vorticity is acted upon by the fluid-deformation tensor and the gradient noise term. 

\section{\bf Integration of the Poisson-de Rham equation} 
\hspace{2em}In $(68)$ we have that $u_\tau$ verifies $(65)$, for every $\tau \ge 0$ 
which from $(28)$ we can rewrite as
\begin{eqnarray}
H_1(g,0)u_\tau = -{1 \over 2}\delta \Omega_\tau, ~{\rm for~any}~\tau \ge 0.
\end{eqnarray}
This last representation together with the Ito-Elworthy formula for $1$-forms, indicates automatically what the representation for the solution is: We have to construct a Jacobian process on $TM$ originated from the derivative of the scalar diffusion with zero drift and diffusion tensor defined by $g$.

Thus, consider the autonomous s.d.e. generated by $H_0(g,0) = {1 \over 2}\triangle _g$:
\begin{eqnarray}
dx^{g,x}_s = X(x^g_s)dW_s, x^{g,x}_0 = x.
\end{eqnarray}
We shall solve the Dirichlet problem in an open set $U$ (of a partition of unity) of $M$ given by $(74)$
with the boundary condition $u_\tau \equiv \phi$ on $\partial U$, with $\phi$ a given $1$-form. Then one can "glue" the solutions
and use the strong Markov property
to obtain a global solution (cf. [31]). Consider the derived velocity process $v^g(s) = (x^g(s),\tilde v^g(s))$ on $TM$, with $\tilde v^g(s) \in T_{x^g(s)}M$, 
whose i.g. is $H_1(g,0)$, i.e. from $(35)$ we have:  
\begin{eqnarray}
d\tilde v^{g,v(x)}_s (x^{g,x}_s) = \nabla^g X(x^{g,x}(s))(\tilde v^{g,v(x)}(s))dW(s),
\end{eqnarray}
with initial velocity $\tilde v^{g,v(x)}(0) = v(x)$.
Notice that equations $(75 \& 76)$ are obtained by taking $u \equiv 0$ in equations $(69 \& 71)$ respectively,
and further rescaling by $(2\nu)^{-{1 \over 2}}$.
Then if $u_\tau$ is a solution of $(74)$ for any fixed $\tau \in [0,T]$, applying to it the
Ito-Elworthy formula and assuming further that
$\delta \Omega_\tau$ is bounded, we then obtain that the formal $C^{1,2}$ solution of the Dirichlet
problem is given by:
\begin{eqnarray}
\tilde u_\tau (x)(v(x))= E^B_x[\phi (x^{g,x}_{\tau_e})(v^{g,v(x)}_{\tau_e} + \int^{\tau_e}_0
{1 \over 2}\delta \Omega_\tau(x^{g,x}_s(v^{g,v(x)}_s)ds]\nonumber
\end{eqnarray}
\begin{eqnarray}
=  \int [\phi(x^{g,x}_{\tau_e})(v^{g,v(x)_{\tau_e}})
+  1/2\int_0 ^{\tau_e}\delta\Omega_\tau (y)(v^{g,v(x)}_s(y)ds]p^g(s,x,y)vol_g(y),
\end{eqnarray}
where $\tau_e = {\rm inf} \{s: x^{g,x}_s \notin U\}$, the first-exit time of $U$ of the process
$\{x^{g,x}_s\}$, and $E^B$ denotes the expectation value with respect to $p^g(s,x,y)$, the transition density of the s.d.e. $(76)$, i.e.the fundamental solution of
the heat equation on $M$: 
\begin{eqnarray}
\partial_\tau p(y) = H_0(g,0)(y)p(y) \equiv 1/2 \triangle_g p(y)
\end{eqnarray}
with $p(s,x,-) = \delta_x$ as $s \downarrow 0$. 

{\bf Theorem~9:} Assume $g$ is uniformly
elliptic, and $U$ has a $C^{2,\epsilon}$-boundary, and furthermore $g^{\alpha \beta}$ and
$\delta \Omega_\tau$
are Hoelder-continuous of order $\epsilon$ on $U$ and $u_\tau$ is uniformly Hoelder-continuous
of order $\epsilon$, for $\tau \in [0,T]$. Then the solution of the Dirichlet problem above
has a unique solution belonging to $C^{2,\epsilon}(U)$ for each $\tau \in [0,T]$ (the maximum principle) [31,47]. Assume instead
that $u_\tau \in H^1(T^*U)$ for each $\tau \in [0,T]$, i.e. belongs to the Sobolev space of order $1$. If $\delta\Omega_\tau \in H^{k-1}(\Lambda^1(T^*U))$,
then $u_\tau \in H^{k+1}(\Lambda^1 (T^*U))$, for $k\ge 1$ and $\tau \in [0,T]$ (cf. [53]).

Notice that in the representations $(73 \& 77)$, 
the local dependance on the curvature 
is built-in (the curvature is defined by second-order derivatives).
This dependance might be exhibited through
the scalar curvature term in the Onsager-Machlup lagrangian
appearing in the path-integral representation of the fundamental solution of the transition
densities of equations $(73)$ and $(77)$ [35,44]. There is further a dependance of the solution on 
the global geometry and topology of $M$ appearing through the Riemannian
spectral invariants of $M$ in the short-time asymptotics of these transition densities [28,29,43].

\section{Kinematic Dynamo Problem of Magnetohydrodynamics}

\hspace{2em}The kinematic dynamo equation for a passive magnetic field transported by an incompressible fluid, is the system of equations [56] for the
time-dependant magnetic vectorfield $B(\tau,x)= B_\tau(x)$ on $M$ defined by $i_{B_\tau}\mu (x) 
= \omega_\tau(x)$ (for $\tau \geq 0$), satisfying
\begin{equation} \label{eqn}
\partial_\tau \omega + (L_{\hat u_\tau} - \nu^m\triangle_{n-1})\omega_t=0, \omega(0,x) = \omega(x), 0 \leq t,
\end{equation}
where $\nu^m$ is the magnetic diffusivity, and we recall that $\mu = vol(g) = det(g)^{1\over 2}dx^1\wedge ... \wedge dx^n$ is the Riemannian
volume density ($(x^1,\ldots,x^n)$ a local coordinate system on $M$), and $\omega \in \Lambda^{n-1}(R\times T^*M)$. In $(79)$, $u$ is assumed given, and it may either be
a solution of NS, or of the Euler equation given by setting $\nu = 0$ in $(63)$. From the definition follows that ${\rm div}B_\tau \equiv 0$, for any $\tau \geq 0$.
Now we note that from $(28)$ we can rewrite this problem as
\begin{eqnarray}
\partial_\tau \omega = H_{n-1}(2\nu^m g, -{1\over 2\nu^m}u_\tau)\omega_\tau, \omega(0,x) = \omega(x), 0 \leq \tau,
\end{eqnarray}
as a linear evolution equation for a $(n-1)$-form, 
similar to the evolution Navier-Stokes equation for the vorticity.
Now if we assume that there is an isometric embedding $f:M \rightarrow R^m$, so that the
diffusion tensor $X = \nabla f$, we can take the Lagrangian representation
for the scalar diffusion generated by $H_0(2\nu^mg, -{1\over 2\nu^m}u_\tau)$, i.e.
the Ito s.d.e. given by substituting $\nu^m$ instead of $\nu$ in equation $(68)$,
and we consider the jacobian process given by $(69-72)$ and
with $\nu^m$ instead of $\nu$,
then the formal $C^{1,2}$ solution of $(80)$ defined on $[0,T]\times M$ for some $T >0$, is given by
\begin{eqnarray}
\tilde \omega_\tau (\Lambda^{n-1}v(x))= E_x[\omega_0(x^{\nu^m,\tau ,x}_\tau)(\Lambda^{n-1}\tilde v^{\nu^m ,\tau ,v(x)}_\tau)]
\end{eqnarray}

{\bf Remarks~8:} We note that similarly to the representation for the vorticity, instead
of the initial vorticity, now it is the initial magnetic form which is transported backwards along the
scalar diffusion, where now the parameter is the magnetic diffusivity, and along its way it is deformed by the fluid-deformation tensor
and the gradient diffusion tensor noise term (this accurately represents the actual
macroscopical physical phenomena yet in a microscopic approach), and finally we take the average for all those 
paths starting at $x$. For both equations as well as the Poisson-de Rham equation,
we have a microspic description which clearly evokes the Feynman approach to Quantum
Mechanics through a summation of the classical action of the mechanical
system along non-differentiable paths. In distinction with the usual Feynman approach,
these Brownian integrals are well defined and they additionally have a clear physical
interpretation which coincides with actual experience.   

\section{Random Implicit Integration Of The Navier-Stokes Equations For Compact Manifolds}
\hspace{2em}Up to this point, all our constructions have stemmed from the
fact that for gradient diffusion processes, the Ito-Elworthy formula shows that the random
process on $\Lambda^2TM$ given by 
$\{\Lambda^2v_\tau: \tau \ge 0\}$ with $\{v_\tau:\tau \ge 0\}$ the jacobian process
fibered on the diffusion
process $\{\Lambda^0 v_\tau \equiv x_\tau: \tau \ge 0\}$ on $M$ given by $(68)$
, is a random Lagrangian flow for the Navier-Stokes equation. Our previous constructions 
have depended on the form of the isometric embedding of $M$. This construction is
very general, since that from a well known theorem due to J. Nash (1951), such an immersion
exists of class $C^1$ for any smooth manifold (cf. [53]). (Furthermore, our assumption of compactness
is for the obtention of a random flow which is defined for all times, and gives
a global diffeomorphism of $M$. The removal of this condition, requires to consider
the random flow up to its explosion time, so that in this case we have a 
local diffeomorphism of $M$.) 

There is an alternative construction of diffusions of differential forms which does not depend on the embedding of $M$
in Euclidean space, being thus the objective of the following section its presentation.
A fortiori, we shall apply these constructions to integrate NS and the kinematic
dynamo problem.

\subsection{The Generalized Hessian Flow}
\hspace{2em} In the following $M$ is a complete compact
orientable smooth manifold without boundary. We shall construct another flow in distinction
of the derived \break flow of the previous sections, which depends explicitly
of the curvature of the manifold, and also of the drift of the diffusion
of scalars. We start by considering an autonomous drift vector field $\hat Q$ (further
below we shall lift this condition) and we define a flow $W^{k,\hat Q}_\tau$ on $\Lambda^kT^*M$ ($1 \le k \le n$) over the flow of $(35)$,
$\{F_\tau(x_0):
\tau \ge 0\}$, by the covariant equation 
\begin{eqnarray}
{D^gW^{k,\hat Q}_\tau \over \partial \tau}(V_0) = -1/2R^k(W^{k,\hat Q}_\tau (V_0)) + (d\Lambda^k)(\nabla^g \hat Q(.))(W_\tau^{k,Q}(V_0)),
\end{eqnarray}
with $V_0 \in \Lambda^kT_{x_0}M$; in this expression the operator ${D^g \over \partial \tau}$ denotes
parallel transport along the curves $x_\tau$ with $\nabla^g$; $R^k$ is the Weitzenbock
term (see [14]) appearing in the Weitzenbock formula for $k$-forms: $\triangle_k = (\nabla^g)^2 - R^k$.
Let $\phi$ be a $k$-form in $C^{1,2}(M)$, and
$V_\tau = W_\tau^{k,\hat Q}(V_0)$ and $x_\tau = F_\tau (x_0)$; then from the Weitzenbock
and general Ito formula we have the following Ito-Elworthy formula [27]:
\begin{eqnarray}
\phi(V_\tau) = \phi (V_0) + \int_0^\tau \nabla^g\phi(X(x_s)dW_s)V_s +
\int_0^\tau [H_k(g,Q)]\phi (V_s)ds;
\end{eqnarray}
In other words, $H_k(g,Q)$ is the i.g. of $\{V_\tau\}$.
In the case that $Q$ is exact the flow $V_\tau$ is called the Hessian
flow. Assume that $1/2R^k -(d\Lambda)^k(\nabla^g \hat Q))(.))$
is bounded below, i.e. for any $V \in \Lambda^k TM$ with $|V| = 1$ we have
\begin{eqnarray}
-\infty < C^k(Q) \equiv inf {1 \over 2}<R^k(V),V> - <(d\Lambda)^k(\nabla^g\hat Q)(.))V,V>,
\end{eqnarray}
where we have denoted by $< , >$ the induced metric on $\Lambda^kTM$. 

{\bf Proposition 2 (Elworthy [27])} Assume that ${1 \over 2}R^k - (d\Lambda)^k(\nabla^g \hat Q)(.))$
is bounded below. Define $P_\tau^k: L^\infty \Lambda^k T^*M \to L^\infty \Lambda^kT^*M$
by 
\begin{eqnarray}
P^k_\tau (\phi)(V) = E(\phi (W_\tau ^{k,Q}(V))
\end{eqnarray}
for $V \in \Lambda^kT_xM, \phi \in L^\infty \Lambda^k T^*M$. Then $\{P^k_\tau: \tau \ge 0\}$ is a contraction semigroup
of bounded continuous forms and is strongly continuous there with i.g. agreeing
with $H_k(g,Q)$ on $C^2(M)$. 

Under the above conditions we can integrate the heat equation for bounded twice
differentiable $k$-forms of class $C^2$
($0 \le k \le n$) and in the general case of a non-autonomous drift vector field $\hat Q = \hat Q_\tau (x)$. Indeed, for every $\tau \ge 0$ consider the flow
$V^\tau_s = W^{k,\hat Q}_{\tau,s}$ over the flow of $\{x^\tau_s: 0 \le s \le \tau\}$,
given by the equation
\begin{eqnarray*}
dx^{x,\tau}_s = X(x^{x,\tau}_s)dW_s + \hat Q_{\tau - s}(x^{x,\tau}_s)ds, x^{x,\tau}_0 = x,
\end{eqnarray*}
obtained by integration of the equation
\begin{eqnarray}
{D^gV^\tau_s \over \partial s}(v_0)=  -1/2R^k(V^\tau_s (v_0)) + (d\Lambda^k)(\nabla^g \hat Q_{\tau-s}(.))(V^\tau_s(v_0)),
\end{eqnarray}
with $v_0 = V^\tau_0 \in T_xM$.
Then, applying the Ito-Elworthy formula we prove as before that if $\tilde \alpha_\tau$ is a bounded
$C^{1,2}$ solution of the Cauchy
problem for the heat equation for $k$ forms:
\begin{eqnarray}
{\partial \over \partial \tau}\alpha_\tau = H_k(g,Q)\alpha_\tau
\end{eqnarray}
with initial condition $\alpha_0(x) = \alpha(x)$ a given $k$-form of class $C^2$,
then the solution of the heat equation is
\begin{eqnarray}
\alpha_\tau(v(x)) = E_x[\alpha_0(V^\tau_\tau(v(x))],
\end{eqnarray}
with $V^\tau_\tau(x)$ the generalized Hessian flow over the flow $\{F_\tau(x): \tau \ge 0\}$
of $\{x^\tau_\tau: \tau \ge 0\}$
with initial
condition $v(x)$.

To integrate the Poisson-de Rham equation we shall need to consider the so-called
Ricci-flow
$W_\tau^{\cal R} \equiv W_\tau^{1,0} (\omega):TM \to TM$ over the random flow generated
by $H_0(g,0)$, obtained by integrating the covariant equation (so we fix the drift to
zero and further take $k=1$ in $(82)$)
\begin{eqnarray}
{D^g W_\tau^{\cal R} \over \partial \tau}(v_0) = -{1 \over 2}\tilde Ric(W_\tau^{\cal R}
(v_0),-), v_0 \in T_{x_0}M
\end{eqnarray}
where $Ric:TM \oplus TM \to R$ is the Ricci curvature and $\tilde Ric (v,-) \in T_xM$
is the conjugate vector field defined by $<\tilde Ric (v,-),w> = Ric(v,w)$, $w \in T_xM$.

\subsection{Integration of the Cauchy problem for the Vorticity on Compact Manifolds}
\hspace{2em} 
{\bf Theorem 10:} The integration of the equation $(67)$ with initial condition $\Omega(0,)=
\Omega_0$
yields
\begin{eqnarray}
\Omega_\tau (v(x)) = E_x[\Omega_0(V^\tau_\tau (v(x))]
\end{eqnarray}
where $\{V^\tau_\tau: \tau \ge 0\}$ is the solution flow over the flow of $\{x^{\nu,\tau,x}_\tau: \tau \ge 0\}$ (see equation $(69)$) of the covariant equation
\begin{eqnarray}
{D^g W^{2,-\hat u_0} \over \partial \tau}(v(x)) & = & -\nu R^2(W^{k,-\hat u_0}_\tau (v(x)))\nonumber\\
& -&  (d\Lambda^2)(\nabla^g \hat u_0(.))(W_\tau^{2,-\hat u_0}(v(x)))
\end{eqnarray}
with initial condition $v(x)\in T_xM$. In this expression, $\nabla^g \hat u_0(.)$
is a linear transformation, $A$, between $T^*_xM$ and $T_xM$, and $d\Lambda^2 (A): T_xM \wedge
T_xM \rightarrow T_xM \wedge T_xM$ is given by $d\Lambda^2 A(v_1\wedge v_2)= Av_1\wedge v_2+
v_1\wedge Av_2$, for any $v_1,v_2 \in T_xM$, $x\in M$.

\subsection{Integration of the Kinematic Dynamo for Compact Manifolds}

Substituting the magnetic diffusivity $\nu^m$ instead of the kinematic viscosity
in $(69)$ and we further consider 
$\{V^\tau_s: s\in [0,\tau]\}$ given by the solution flow over the flow of 
$\{x^{\nu^m,\tau,x}_s: s \in [0,\tau]\}$ (see equation $(69)$) of the covariant equation
\begin{eqnarray}
{D^gV^\tau_s \over \partial s}(v(x))=  -\nu^mR^{n-1}(V^\tau_s (v(x))) + (d\Lambda^{n-1})(\nabla^g \hat Q_{\tau-s}(.))(V^\tau_s(v(x))),
\end{eqnarray}
with $v(x) = V^\tau_0 \in T_xM$. Then, the formal $C^{1,2}$ solution of $(80)$ is
\begin{eqnarray}
\omega_\tau (v(x)) = E_x[\omega_0(V^\tau_\tau (v(x))].
\end{eqnarray}
with $E_x$ denoting the expectation valued with respect to the measure on $\{x^{\nu^m,\tau,x}_\tau\}$
(whenever it exists).

\subsection{Integration of the Poisson-de Rham Equation for the Velocity}

With the same notations as in the case of isometrically embedded manifolds, we
have a martingale problem with a bounded solution given by
\begin{eqnarray}
u_\tau (v(x)) = E^B_x[\phi(W^{{\cal R}}_{\tau_e}(v(x))) + 1/2\int^{\tau_e}_0\delta\Omega_\tau(W^{{\cal R}}_s(v(x)))ds]
\end{eqnarray}

\section{Representations Of NS On Euclidean Space}
\hspace{2em}

In the case that $M$ is euclidean space, the solution of NS
is easily obtained from the solution in the general case. In this case the isometric embedding $f$
of $M$ is realized by the identity mapping, i.e. $f(x) = x, \forall x \in M$.
Hence the diffusion tensor $X =I$, so that the metric $g$ is also the identity.
For this case we shall assume that the velocity vanishes at infinity, i.e.
$u_t \rightarrow 0$ as $|x| \rightarrow \infty$. (This allows us to carry
out the application of the general solution, in spite of the non-compacity
of space). Furthermore, $\tau_e = \infty$.
The solution for the vorticity equation results as follows. We have the s.d.e. (see $(69)$
where we omit the kinematical viscosity, for simplicity)
\begin{eqnarray}
dx^{\tau,x}_s = -u(\tau -s,x^{\tau,x}_s)ds + (2\nu)^{1\over 2}dW_s,x^{\tau,x}_0 = x,s \in [0,\tau].
\end{eqnarray}
The derived process is given by the solution of the o.d.e. (since in $(71)$ we have $\nabla X \equiv 0$)
\begin{eqnarray}
d\tilde v^{\tau,x,v(x)}_s = -\nabla u(\tau-s,x^{\tau,x}_s)(\tilde v^{\tau,x,v(x)}_s)ds, v^{\tau,v(x)}_0 = v(x)\in R^n, s\in [0,\tau],
\end{eqnarray}

Now for $n= 3$ we have that the vorticity $\Omega(\tau,x)$ is a $2$-form on $R^3$, or still by
duality has an adjoint $1$-form, or still a function, which with abuse of notation
we still write as $\tilde \Omega (\tau,.):R^3\rightarrow R^3$, which from $(73)$ we can write as
\begin{eqnarray}
\tilde \Omega(\tau,x) = E_x[\tilde v^{\tau,x,I}_\tau \Omega_0(x^{\tau,x}_\tau)],
\end{eqnarray} 
where $E_x$ denotes the expectation value with respect to the measure (if it exists)
on $\{x^{\tau,x}_\tau:\tau \ge 0\}$,
for all $x \in R^3$, and in the r.h.s. of $(97)$ we have matrix multiplication Thus, in this case, we have that the deformation tensor
acts on the initial vorticity along the random paths. This action is the one
that for $3D$ might produce the singularity of the solution. Note that in Euclidean
space there is no gradient-noise contribution to the folding of the initial
vorticity by the action of the fluid deformation tensor.

In the case of $R^2$, the vorticity can be thought as a pseudoscalar, since $\Omega_\tau (x)=\tilde \Omega_\tau(x)dx^1\wedge dx^2$,
with $\tilde \Omega_\tau :R^2\rightarrow R$, and being the curvature identically
equal to zero, the vorticity equation is (a ${\it scalar}$ diffusion equation)
\begin{eqnarray}
{\partial\tilde \Omega_\tau \over \partial \tau}= H_0(2\nu I,{-1 \over 2\nu}u_\tau)\tilde \Omega_\tau
\end{eqnarray}
so that for $\tilde \Omega_0 = \tilde \Omega$ given, the solution of the initial 
value problem is
\begin{eqnarray}
\tilde \Omega(\tau,x) = E_x[\tilde \Omega (x^{\tau,x}_\tau)]
\end{eqnarray}
This solution is qualitatively different from the previous case. Due
to a geometrical duality argument, for $2D$ we have factored out completely the derived process
in which the action of the deformation tensor on the initial vorticity is present.

Furthermore, the solution of equation $(75)$ is (recall that $X= I$)
\begin{eqnarray}
x^{g,x}_\tau = x + W_\tau,
\end{eqnarray}
and since $\nabla X = 0$, the derived process (see $(76)$) is constant 
\begin{eqnarray}
v^{g,x,v(x)}_\tau = v(x),\forall \tau \in [0,T].
\end{eqnarray}
so that its influence on the velocity of the fluid can be factored out in the
representation $(77)$. Indeed, we have
\begin{eqnarray*}
\tilde u_\tau (x)(v(x)) & = & E^B_x[\int^\infty _0{1 \over 2}\delta \Omega_\tau(x + W_s)(\tilde v^{g,x,v(x)}(s))ds]\nonumber\\
& =& E^B_x[\int^\infty_0 {1\over 2}\delta \Omega_\tau(x +W_s)ds(v(x))]
\end{eqnarray*}
for any tangent vector $v(x)$ at $x$, and in particular (we take $v(x) = I$) we obtain
\begin{eqnarray}
\tilde u_\tau (x) = E^B_x[\int^\infty _0{1 \over 2}\delta \Omega_\tau(x + W_s)ds].
\end{eqnarray}
In this expression we know from $(78)$ that the expectation value is taken with respect to
the standard Gaussian function, $p^g(s,x,y) = (4\pi s)^{-n\over 2}exp(-{|x-y|^2 \over 4s})$.

Let us describe in further detail this solution separately for each dimension. We note first
that if $\Omega_\tau \in L^1\cap C^1_b$ (where $C^1_b$ means continuously differentiable, bounded
with bounded derivatives) 
\begin{eqnarray}
E[\delta \Omega_\tau (x+ W_s)] = \delta E[\Omega_\tau (x+W_s)]
\end{eqnarray}
In case $n=2$, for a $2$-form $\beta$ on $M$ we have $\delta \beta = \delta (\tilde \beta dx^1\wedge dx^2)
=- (\partial _2 \tilde \beta dx^1 -\partial _1 \tilde \beta dx^2) \equiv -\nabla^\perp \beta$.
In case $n= 3$, for a vorticity described by the $1$-form (or a vector-valued function) $\tilde \Omega_\tau : R^3\rightarrow R^3$
adjoint to the vorticity $2$-form $\Omega_\tau$, we have that
\begin{eqnarray}
\delta \Omega_\tau =- d\tilde \Omega_\tau = -{\rm rot} \tilde \Omega_\tau.
\end{eqnarray}
Therefore, we have the following expressions for the velocity: When $ n=2$ we have
\begin{eqnarray}
u_\tau (x) = \int ^\infty_0 -{1\over 2}\nabla^\perp E^B_x[\tilde \Omega_\tau (x+W_s)]ds
\end{eqnarray}
while for $n= 3$ we have 
\begin{eqnarray}
u_\tau(x) = \int^{\tau_e}_0 {-1\over 2}dE^B_x[\tilde \Omega_\tau (x+W_s)]ds.
\end{eqnarray}

Now we can obtain an expression for the velocity which has no derivatives
of the vorticity; this follows the basic idea in a non-geometrical
construction given by Busnello who starts
with the stream function (of an unbounded incompressible fluid) instead of the velocity [54]. Consider the semigroup generated by $H_0(I,0)= {1\over 2}\triangle$,
i.e. $P_s \tilde \Omega_\tau (x) = E[\tilde \Omega _\tau (x+W_s)]$ (in the case $n=3$
this means the semigroup given on each component of $\tilde \Omega$). From the Elworthy-Bismut
formula valid for scalar fields (see [52]) we have that (in the following $e_i,i=1,2,3$ denotes the canonical base in $R^2$
or $R^3$)
\begin{eqnarray*}
\partial _i P_s \tilde \Omega_\tau(x) = <dP_s\tilde \Omega (x),e_i> = {1\over s}E^B_x[\tilde \Omega (x+W_s)\int^s_0 <e_i,dW_r>] 
\end{eqnarray*}
\begin{eqnarray}
= {1\over s}E^B_x[\tilde \Omega (x+W_s)\int^s_0 dW_r^i]
= {1\over s}E[\tilde \Omega_\tau (x+W_s)W^i_s].
\end{eqnarray}
Therefore, for $n=2$ we have from $(105,107)$
\begin{eqnarray}
u_\tau(x) = - \int ^{\tau_e}_0 {1\over 2s}E^B_x[\tilde \Omega_\tau (x+W_s)W^\perp_s]ds
\end{eqnarray}
where $W^\perp_s = (W^1_s,W^2_s)^\perp = (W^2_s,-W^1_s)$. Instead, for $n=3$ we have from $(106,107)$ that
\begin{eqnarray}
u_\tau(x) = -\int^{\tau_e}_0 {1\over 2s}E^B_x[\tilde \Omega_\tau (x+W_s)\times W_s]ds
\end{eqnarray}
where $\times$ denotes the vector product and $W= (W^1,W^2,W^3) \in R^3$. 
 
Thus, we have obtained the representations for NS in 2D and 3D:

\subsection{Integration of the Kinematic Dynamo Problem in Euclidean space}
With the notations in this section, the kinematic dynamo problem
in $3D$ can be solved as follows. As for the vorticity, the magnetic field is
for $n= 3$ is a $2$-form on $R^3$, or still by
duality has an adjoint $1$-form (so the argument turns to work out as well for 2D), or still a function, which with abuse of notation
we still write as $\tilde \omega (\tau,.):R^3\rightarrow R^3$, which from $(81)$ we can write as
\begin{eqnarray}
\tilde \omega(\tau,x) = E_x[\tilde v^{\tau,x,I}_\tau \omega_0(x^{\tau,x}_\tau)],
\end{eqnarray} 
where $E_x$ denotes the expectation value with respect to the measure (if it exists)
on $\{x^{\tau,x}_\tau:\tau \ge 0\}$,
for all $x \in R^3$, and in the r.h.s. of $(110)$ we have matrix multiplication Thus, in this case, we have that the deformation tensor
acts on the initial vorticity along the random paths. This action is the one
that for $3D$ produces the complicated topology of transported magnetic fields.
This solution was obtained independently by Molchanov et al [50] and further applied in
numerical simulations (see Ghill and Childress [51] and references therein).

\section{The Navier-Stokes Equation is Purely Diffusive For Any Dimension
Other than $1$}
\subsection{Motivations}
\hspace{2em} We have given up to now a derivation of diffusion processes
starting from gauge theoretical structures, and applied this to give implicit representations for the covariant Navier-Stokes equations. These constructions were possible as they stemmed from the extremely tight relation existing between the metric-compatible Riemann-Cartan-Weyl connections, and the diffusion processes for differential forms, built untop of the diffusions for scalar fields. As we saw already this stemmed from the fact that there is a one-to-one
correspondance between said RCW connections and the scalar diffusion processes
$\{x_\tau: \tau \ge 0\}$ with drift given by $\hat Q$ and diffusion tensor $X$. 
As we can easily check from $(34)$, this construction is valid for $n \neq 1$. This leads to conjecture that in a gauge theoretical setting and further applying stochastic analysis, one could do away with the drift, in any dimension other than $1$. If this would be the case, then we could apply this construction to the Navier-
Stokes equation, which thus in any dimension other than $1$ would turn to be
representable by random lagrangian paths which do not depend explicitly on the velocity of the fluid, since they would be purely diffusive processes. 

 \subsection{More On Connections}
\hspace{2em}
Consider the map $M\times R^ m \rightarrow TM \rightarrow 0$ which we assume that it has a right inverse $Y:TM \rightarrow M \times R^ m$. Here, $Y =X^ \dagger$ is the adjoint of $X$ with respect to the Riemannian metric on $TM$ induced by $X$, $Y = X^*$. Write $X( x) = X(x,.):R^ m \rightarrow TM$. For $u \in TM$, let $Z^ u \in \Gamma(TM)$ defined by
\begin{equation}
Z^ u(x) = X(x)Y(\pi (u))u.
\end{equation}

{\bf Proposition 3:} There is a unique linear connection $\tilde \nabla$ on $TM$ such that for all $u_0 \in T_{x_0}M, x \in M$, we have that
\begin{equation}
\tilde \nabla_{v_0}Z^ {u_0} = 0.
\end{equation}
It is the pushforward connection defined as
\begin{equation}
\tilde \nabla _{v_0}Z := X(x_0)d(Y(.)Z(.))(v_0), v_0 \in T_{x_0}M, Z \in \Gamma(TM),
\end{equation}
where $d$ is the usual derivative of the map $Y(.)Z(.):M \rightarrow R^ m$.

{\bf Proof:} The above definition defines a connection. Let $\hat\nabla$ be any linear 
connection on $TM$. We have
\begin{equation}
Z(.) =X(.)Y(.)Z(.).
\end{equation}
Then, for $v \in T_{x_0}M$, 
\begin{equation}
\hat\nabla _v Z = X(x_0)d(Y(.)Z(.))(v) + \hat\nabla_v [X(.)(Y(x_0)Z(x_0))]
= \tilde \nabla _vZ + \hat\nabla _v Z^ {Z(x_0)}.
\end{equation}
Since $\hat\nabla$ is a connection by assumption, and since the map
\begin{equation}
TM \times TM \rightarrow TM, (v,u) \longmapsto \hat\nabla _v Z
\end{equation}
gives a smooth section of the bundle $Bil(TM\times TM;TM)$, then $\hat\nabla$
is a smooth connection on $TM$. Taking $\hat\nabla = \tilde \nabla$ we obtain a connection with the desired property. 

{\bf Theorem 11:} Let $Y$ be the adjoint of $X$ with respect to the induced metric on $TM$ by $X$. Then, $\tilde \nabla$ is metric compatible, where the metric is the one induced by $X$ on $TM$, which we denote by $\tilde g$. Moreover, since $M$ is finite-dimensional, any metric-compatible connection for any metric on $TM$ can be obtained this way from such $X$ and $R^ m$.

{\bf Proof:}  We have
\begin{displaymath}
2\tilde g(\nabla_vU,U) =  2(X(x_0)(d(Y(.)U(.))(v),U(x_0))\nonumber\\
\end{displaymath}
\begin{equation}
 =  2\tilde g(d(Y(.)U(.))(v),Y(x_0)U(x_0)) =d(\tilde g(U,U)(v),
\end{equation}
so that $\tilde \nabla$ is indeed metric compatible. By the Narasimhan-Ramanan theorem on universal connections [51], any metric compatible connection arises likes this. Indeed, $\tilde \nabla$ is the pull-back of the universal connection over the Grassmanian $G(m,n)$ of $n$-planes in $R^ m$ by the map $x \mapsto [{\rm Image} Y(x):T_xM \rightarrow R^m]$. In particular, the RCW
connections arise from such a construction. c.q.d.

Two connections, $\nabla^a$ and $\nabla^b$ on $TM$ give rise to  a bilinear map $D^{ab}:TM \times TM \rightarrow TM$ such that
\begin{equation}
\nabla^a _VU = \nabla^b_VU + D^{ab}(V,U), U,V \in \Gamma (TM).
\end{equation}
 Choose $\nabla^ b = \nabla^ g$, the Levi-Civita connection of a certain Riemannian metric $g$. Consider
\begin{equation}
\tilde \nabla _vU = \nabla^ g_VU + \tilde D(V,U),
\end{equation}
where we decompose $\tilde D$ into
\begin{equation}
\tilde D(u,v) = A(u,v) + S(u,v),
\end{equation}
where
\begin{equation}
A(u,v) = -A(v,u), S(u,v) = S(v,u).
\end{equation}
Since the torsion tensors $T^a$ and $T^b$ of any two connections $\nabla^a$
and $\nabla^b$ respectively, are connected through the expression
\begin{equation}
T^a(u,v) + T^b(u,v) = D^{ab}(u,v) -D^{ab}(v,u).
\end{equation}
which for the case of $\nabla^ b = \nabla^ g$ as $T^b = 0$, we can write for
$D^{ab} = \tilde D$, the identity
\begin{equation}
\tilde T(u,v) = \tilde D(u,v) - \tilde D(v,u) + 2A(u,v).
\end{equation}
Thus, 
\begin{equation}
A(u,v) = {1 \over 2}\tilde T(u,v), u,v, \in \Gamma(TM), \tilde T \equiv T^a.
\end{equation}
Therefore, the decomposition in $(119)$ is written in the form
\begin{equation}
\tilde \nabla _vU = \nabla^ g _vU + {1\over 2}\tilde T(u,v) + S(u,v), u,v \in \Gamma(TM).
\end{equation}
which is nothing else than the original decomposition for a metric compatible Cartan connection given in $(10 \&11)$.
As we know already from the decomposition $(11)$, we have the following lemma.

{\bf Lemma } A connection $\tilde \nabla$ is metric-compatible if and only if the map $\tilde D(v,.):TM \rightarrow TM$ is skew-symmetric for each $v \in TM$, i.e.
\begin{equation}
g(\tilde D(v,u_1),u_2) + g(\tilde D(v,u_2),u_1) = 0, u_1,u_2 \in \Gamma (TM).
\end{equation}
Equivalently,
\begin{equation}
g(S(u_1,u_2),v) = {1 \over 2}g(\tilde T(v,u_1),u_2) + {1\over 2}g(\tilde T(v,u_2),u_2).
\end{equation}
Consequently, for $U_1$, $U_2$,$V$ in $\Gamma (TM)$, we have
\begin{equation}
g(\tilde D(V,U_1),U_2) = {1\over 2}g(\tilde T(V,U_1),U_2) + {1\over 2}g(\tilde T(U_2,V),U_1) + {1\over 2}g(\tilde T(U_2,U_1),V),
\end{equation}
which is decomposition $(10)$.

{\bf Proof:} Take $V,U_1,U_2 \in \Gamma (TM)$. Then,
\begin{displaymath}
d(g(U_1,U_2))(V) = g(\nabla^ g_VU_1,U_2) + g(U_1,\nabla^ g_VU_2)\nonumber\\
\end{displaymath}
\begin{equation}
=  g(\nabla^ g_VU_1,U_2) + g(U_1,\nabla^ g_VU_2) - g(\tilde D(V,U_1),U_2) - g(U_1,\tilde D(V,U_2).
\end{equation}
So, $\tilde \nabla$ is metric compatible if and only if 
\begin{equation} 
g(\tilde D(V,U_1),U_2) + g(U_1,\tilde D(V,U_2)) = 0.
\end{equation}
Now, writing $\tilde D = A + S$ we get
\begin{equation}
g(A(V,U_1),U_2) + g(A(V,U_2),U_1) = -g(S(V,U_1),U_2) - g(S(V,U_2),U_1). 
\end{equation}
We now
observe that for an alternating bilinear map $L:TM\times TM \rightarrow TM$,
\begin{equation}
Cyl[g(L(v,u_1),u_2) + g(L(v,u_2),u_1)] = 0,
\end{equation}
where $Cyl$ denotes cyclic sum. Taking the cyclic sum in equation $(131)$ and apply $(132)$ to $A$, we thus obtain ${\rm Cyl}g(S(V,U_1),U_2) = 0$ which on further substituting in $(131)$ we obtain 
\begin{equation}
g(A(V,U_1),U_2) + g(A(V,U_2),U_1) = g(S(U_1,U_2),V).
\end{equation}

\subsection{The Trace-Torsion Is Dynamically Redundant in Any Dimension Other Than $1$}

Let us return to our original setting of Section $1$. We assume a metric-compatible Cartan connection, which we now write as $\tilde \nabla$
with torsion tensor $\tilde T$. The following result is a reduction of a more general result due to Elworthy, Le Jan and Li {61].

{\bf Theorem 12:} Assume $M$ has dimension bigger than $1$. Consider the laplacian on $0$-forms $H_0(g,Q)$ where $Q$ is the trace-torsion $1$-form
of $\tilde \nabla$, 
\begin{equation}
Q(u) = {\rm trace~} g(\tilde T(-,u),-).
\end{equation}
Assume further that we can write the laplacian $H_p(g,Q)$ on $p$-forms ($0 \le p \le n$) in the Hormander form:
\begin{equation}
{1\over 2}\sum_{i=1}^m L_{V_i}L_{V_i}+ L_Z
\end{equation}
where $Z$ is a vectorfield on $M$, $V:M\times R^ m \rightarrow TM$ is a smooth surjection, linear in the second variable, and $V_i$ is defined by 
\begin{equation}
V(x,e) = V(x)e = \sum_{i=1}^m V^i(x)<e,e_i>, 
\end{equation}
whith $e_1, \ldots, e_m$ the standard orthonormal basis for $R^ m$.
(Since $\nabla^ g$ is metric compatible, from Theorem $11$ and the transformation rules between Stratonovich and Ito calculi, we can always introduce a defining 
map $V$ for $\nabla^ g$ that gives such decomposition (c.f. [27])).
Then, there exists a map $K:M\times R^ m \rightarrow TM$ linear on the second variable, such that the solution to the Stratonovich equation
\begin{equation}
dx_\tau = K(x_\tau)\circ dW_\tau,
\end{equation}
has $H_0(g,Q)$ for infinitesimal generator, i.e.
\begin{equation}
H_0(g,Q) = {1 \over 2}\sum_{i=1}^ m L_{K_i}L_{K_i}.
\end{equation}
In other words, the Ito s.d.e. given by $(35)$ admits a driftless
representation given by $(137)$.

{\bf Proof:} Set for the original drift vectorfield $\hat Q$ (the $g$-conjugate of $Q$), the decomposition
\begin{equation}
\hat Q = {1 \over 2}\sum_{i=1}^ m \nabla^ g_{V^ i}V^ i -Z,
\end{equation}
A connection $\tilde \nabla$ suitable for this is such that
\begin{equation}
2A(u,v)= \tilde T(v,u) = {2 \over n-1}(u\wedge v)Q(x).
\end{equation}
Consider a bundle map $K:M\times R^ m \rightarrow TM$ which gives rise to the metric compatible connection $\tilde \nabla$ (theorem $11$).
Consider the s.d.e. 
\begin{equation}
dx_\tau = K(x_\tau)\circ dW_\tau.
\end{equation}
Its generator is (c.f. [27])
\begin{equation}
{1\over 2}{\rm trace}(\nabla^ g)^ 2 + {1\over 2}\sum_{i=1}^ m \nabla^ gK^i(K^ i)
\end{equation}
while by assumption we have
\begin{equation}
H_0(g,Q) = {1\over 2}{\rm trace }(\nabla^ g)^ 2 + ({1 \over 2}\sum_{i=1}^ m \nabla^ g_{V^ i}V^ i -Z) = { 1 \over 2}{\rm trace}(\nabla^ g)^ 2 + \hat Q
\end{equation}
The required result follows after we show 
\begin{equation}
\sum_{i=1}^ m \nabla^ gK^i(K^i) = -\sum_{i=1}^ m\tilde D(K^i,K^i) = {\rm trace}\tilde D(-,-)
\end{equation}
equals $2\hat Q$. For this we note that for all $v \in TM$,
\begin{equation}
g(\sum_{i=1}^ m \tilde D(K^i,K^i),v) ) = g(\sum_{i=1}^ m S(K^i,K^ i),v)) 
= -\sum_{i=1}^m g(\tilde T(v,K^i),K^ i) \break
= 2g(\hat Q,v)
\end{equation}
Consequently
\begin{equation}
{\rm trace} (\nabla^ g)^ 2 + {1 \over 2}\sum_{i=1}^ m\nabla^ g(K^ i)(K^ i) =
{\rm trace}(\nabla^ g)^ 2 + \hat Q = H_0(g,Q)
\end{equation}
and the $K$ so constructed is the required map.c.q.d.

{\bf Remarks 8:} Of course in the above construction, it is innecessary to start with an arbitrary metric-compatible Cartan connection, only the trace-torsion as proved already matters.

\subsection{Navier-Stokes Equations Is Purely Diffusive In Any Dimension
Other Than $1$}

We recall that for any dimension other than $1$, the Navier-Stokes equation is a diffusion process which arises from
a RCW connection of the form $(34)$ with metric given by $2\nu g$, where $g$ is the original metric
defined on $TM$, and torsion restricted to the trace-torsion given by ${-1\over 2\nu}u_\tau$, where
$\tau \ge 0$: we shall call this connection the Navier-Stokes connection with parameter $\nu$, which we shall denote as $\nabla ^{NS;\mu}$. Let us then consider a bundle map $K_\tau:M\times R^ m \rightarrow TM$, for $\tau \ge 0$ for such a connection; from theorem $11$
we know it exists. Therefore,
from Theorem $12$ we conclude that:

{\bf Theorem 13:} For any dimension other than $1$, the random lagrangian representations given by $(68)$ admit representations as a Stratonovich s.d.e. without drift $-\hat u_\tau$
term
\begin{equation}
dx_\tau = K_\tau(x(\tau))\circ dW_\tau.
\end{equation}

{\bf Remarks 9:} Thus, we have gauged out the velocity in the dynamical representation for
the fluid particles. Of course, the new diffusion tensor $K_\tau$ ($\tau \ge 0$)
depends implicitly in the velocity of the fluid as well as in the kinematical viscosity. Indeed, $K$ can be in principle computed
from the knowledge of the Navier-Stokes connection with parameter $\nu$, by solving the equation
\begin{equation}
\nabla^{NS;\mu}_{v}Z = K(x)d(K^\dagger(.)Z(.)(v), v \in T_xM, Z \in \Gamma (TM).
\end{equation}

{\bf Remarks 10:} This last theorem deserves further examination. As well known, the Navier-Stokes equations is a mechanism of "competition" between the linear diffusion term $\nu \triangle_1 u_\tau$ which at the level of the fluid flow described by the diffusion tensor term $(2\nu)^{1 \over 2}X$ and the drift non-linear term $PL_{u_\tau}u_\tau$, described at the level of the fluid flow by the drift vectorfield $-\hat u_\tau$. The long-time existance of the representations of Navier-Stokes depends on the prevailance of the diffusion term, so that the non-linearity
which feeds the fluid with energy would fade out. The representation given by Theorem $13$ assures that this is the case for any physically interesting dimension, without distinction between 2-dimensional and 3-dimensional manifolds, since we can always find a non-linear representation for the random fluid flow which incorporates
both the diffusion tensor and the velocity into a new diffusion tensor! (We can think here in the dispute on pre-Socratic Greek philosophy
between the followers of Parmenides and Heraclitus, and thus name $(147)$ as the Parmenidean representation: in this representation the fluid flow does not depend explicitly on the fluid velocity.)

{\bf Remarks 11:} Just like in theorem $13$, we can represent the lagrangian random paths for the kinematic dynamo problem as a purely diffusive process, by putting $\nu^ m$ instead of $\nu$.

\section{Final~Observations:}
\hspace{2em} The method of integration applied in the previous section
is the extension to differential forms of the  
method of integration (the so-called martingale problems) of elliptic and parabolic partial differential equations 
for scalar fields [24,31]. In distinction with the Reynolds approach in Fluid Mechanics,
which has the feature of being non-covariant, in the present
approach, the invariance by the group of space-diffeomorphims has been the key
to integrate the equations, in separating covariantly the fluctuations and drift
terms and thus setting the integration in terms of covariant martingale problems.
The role of the RCW connection is precisely to yield this separation for the diffusion
of scalars and differential forms, and thus
the role of the differential geometrical structure is essential. 

The solution scheme we have presented gives rise to infinite particle random trajectories due to the
arbitrariness of the initial point of the Lagrangian paths. This continuous infinite
particle solution is ${\it exact}$ and we have actually
computed its expression. Actually, to integrate NS we choose a finite
set of initial points and we take for $\Omega_0$ a linear combination of $2$-forms
(or area elements in the $2$-dimensional case)
supported in balls centered in these points, the so-called many vortices solutions; one can choose the original $f_0(x)$
so that $\Omega_0$ is supported in these balls and these localizations persists in time.
Thus the role of the potential term in the Buttke magnetization $1$-form in the expression
$(56)$ is to push  the vortices to be confined on predetermined finite radii balls; see Chorin [1].  
This requires that convergence to a solution of NS be proved in addition. 

A new approach to NS as a (${\it random}$) dynamical system appears. Given a 
stationary measure for 
the ${\it random}$ diffeomorphic flow of NS given by the stationary flow of equation $(67)$,
one can construct the state space of this flow and further, its
${\it random}$ Lyapunov spectra. Consequently, assuming ergodicity of this
measure, one can conclude that the moment instability
of the flow is related to a cohomological property of $M$, namely the existance of non-trivial harmonic one-forms
$\phi$, which are preserved by the vectorfield $\hat u$ of class $C^2$, i.e. $L_{\hat u}\phi = di_{\hat u}\phi = 0$; see
page $61$ in [27]. We also have the random flows $\{v_\tau \wedge v_\tau:\tau \ge 0\}$ 
and $W_\tau ^{2,-\hat u_0}$ on $TM\wedge TM$ of Theorem $7$ and Theorem $10$
respectively, which integrate
the linear equation for the vorticity. Concerning these flows, the stability theory
of NS $(67)$ requires an invariant measure on a suitable subspace of $TM\wedge TM$ and 
further, the knowledge of the spectrum of the one-parameter family of ${\it linear}$
operators depending on $\nu$, $H_2(2\nu g,-{1 \over 2 \nu}u_\tau)$.
The latter may play the role of the Schroedinger operators in Ruelle's theory of turbulence,
which were introduced by linearising NS for the velocity as the starting point
for the discussion of the instability theory; see article in pages $295-310$ in Ruelle [48]. 

Finally, it has been established
numerically that turbulent fluids resemble the random motion of dislocations [4].
In the differential geometric gauge theory of crystal dislocations, the torsion tensor is the dislocation
tensor [12], and our presentation suggests that this analogy might be established rigorously from the
perspective presented here. We would like to remark that the results presented in this article, are part of a more general program of
formulation of gravitation, quantum mechanics and irreversible thermodynamics, in terms of stochastic differential geometry, developed by the author. 

\section{Appendix}
We shall review some basic concepts of the probabilistic, analytical
and geometrical realms.

Let $\{{\cal F}_\tau: \tau \ge 0\}$ be a family of sub $\sigma$-fields of a $\sigma$-field
${\cal F}$ in a probability space $(\Omega,{\cal F},P)$. It is called a ${\bf filtration}$
of sub $\sigma$-fields if it satisfies the following three properties: i)${\cal F}_s \subset
{\cal F}_\tau$ if $s < \tau$; ii)$\cap _{\epsilon > 0}{\cal F}_{\tau + \epsilon}
= {\cal F}_\tau$, and iii) each ${\cal F}_\tau$ contains all null sets of ${\cal F}$

A stochastic process $x_\tau, \tau \in T$, with $T$ a time-set, say $[0,\infty)$, the interval
$[0,T]$ or the real line, is called $({\cal F}_\tau)-{\bf adapted}$ if for each
$\tau, x_\tau$ is ${\cal F}_\tau$-measurable. ${\bf Predictable~sets}$ are subsets of
$[0,\infty)\times R$, which are elements of the smallest $\sigma$-algebra relative
to which all real ${\cal F}_\tau$-adapted, right-continuous processes with left-hand
limits are measurable in $(\tau,\omega)$. A stochastic process $x:[0,\infty) \mapsto S$, where $S$
is a measurable space with $\sigma$-algebra ${\cal B}$
is called {\bf predictable} if, for any Borel subset ${\cal B} \in S$, $\{(\tau,\omega), x(\tau,\omega) \in B\}$
is predictable.

A positive random variable $t$ is called a {\bf stopping~time} (with respect to
the filtration $\{{\cal F}_\tau: \tau \ge 0\}$ if for all $0 \leq \tau, \{t \leq \tau\}
\in {\cal F}_\tau$. This concept is used to indicate the ocurrence of some random
event.

The {\bf conditional~expectation} of a real-valued random variable X with respecto
to a sub-$\sigma$ algebra ${\cal G}$ of ${\cal F}$ is denoted by $E(X|{\cal G})$.

Let $x_\tau$ be a 
$({\cal F}_\tau)$-adapted real-valued process such that for each $\tau, x_\tau$ is integrable.
It is called a ${\bf martingale}$ if it satisfies: $E[x_\tau|{\cal F}_s] = x_s$ a.s. for
any $\tau > s$. Furthermore, $x_\tau$ is called a ${\bf local~martingale}$ if there exists
an increasing sequence of stopping times $\{\tau_n\}$ such that $P(\tau_n < T) \to 0$
as $n \to \infty$, and each stopped time $x^{\tau_n}_\tau \equiv x_{\tau \wedge \tau_n}$
is a martingale, where $\tau \wedge \tau_n = min \{\tau,\tau_n\}$. A martingale
is obviously a local martingale (set $\tau_n \equiv T$, for all $n$. Finally, $x_\tau$
is called a ${\bf semimartingale}$ if it can be decomposed as the sum of a localmartingale
and a process of bounded variation.

This decomposition, which in the course of the presentation of the rules
of stochastic analysis on manifolds for differential forms appears 
explicitly in the Ito-Elworthy Formula,
sets the integration of Navier-Stokes equations and the kinematic dynamo problem
as the solutions of the so-called ${\bf martingale~
problems}$ (after Stroock and Varadhan $[30]$). 

Let $D$ be a domain in Euclidean space $R^d$ and let $R^l$ another Euclidean space
(eventually $d = l$). Let $m \in N$; denote $C^m \equiv C^m(D;R^l)$ the set of
all maps $f:D \to R^l$ which are $m$-times continuously differentiable. For the
multi-index $\alpha = (\alpha_1,\ldots,\alpha_d) \in N^d$, define
\begin{eqnarray}
D^\alpha_x = {\partial ^\alpha \over (\partial x^1)^{\alpha_1}\ldots (\partial x^d)^{\alpha ^d}}, {\rm with} |\alpha|= \sum_{i = 1}^d \alpha_i.
\end{eqnarray}
Let $K$ be a subset of $D$. Set
\begin{eqnarray}
||f||_{m,K}= sup_{x \in K}{f(x) \over (1 + |x|)} + \sum_{1 \leq |\alpha| \leq m} sup_{x \in K}|D^\alpha f(x)|.
\end{eqnarray}
Then $C^m(D;R^l)$ is a Frechet space by seminorms $\{||.  ||_{m;K}:$ K are compacts
in $D \}$. When $K = D$ we write $||.||_{m;K}$ as $||.||_K$. Now let $\delta$ such
that $0<\delta<1$. Denote by $C^{m;\delta}\equiv C^{m;\delta}(D;R^l)$ the set
of all $f \in C^m$ such that $D^\alpha f, |\alpha|= m$ are $\delta$-Holder continuous.
By the seminorms
\begin{eqnarray}
||f||_{m+\delta;K}= ||f||_{m;K} + \sum_{\alpha = m}sup_{x,y \in K, x\neq y}
{D^\alpha f(x) - D^\alpha f(y) \over |x-y|^\delta},
\end{eqnarray}
is a Frechet space, the so-called ${\bf space~of~\delta-Holder~continuous~C^m~mappings}$.
When $D = K$ we write $||. ||_{m+\delta;K}$ as $||. ||_{m+\delta}$. Denote further
as $C^{m;\delta}_b$ the set $\{f \in C^{m;\delta}: ||f||_{m+\delta}<\infty\}.$
A continuous mapping $f(\tau,x)$, $x \in D, \tau \in T\}$ is said to belong to the
class ${\bf C^{m;\delta}}$ if for every $\tau, f(\tau) \equiv f(\tau,.)$ belongs
to $C^{m;\delta}$ and $||f(\tau)||_{m+\delta;K}$ is integrable on $T$ with respect
to $\tau$ in any compact subset $K$. If the set $K$ is replaced by $D$, $f$ is said
to belong to the class ${\bf C^{m;\delta}_b}$. Furthermore, if $||f(\tau)||_{m+\delta}$
is bounded in $\tau$, it is said to belong to the class ${\bf C^{m;\delta}_{ub}}$.

Consider the {\bf canonical~Wiener~space} $\Omega$ of continuous
maps $\omega :R \to R^d, \omega (0) = 0$,
with the canonical realization of the Wiener process $W(\tau)(\omega) = \omega (\tau)$. Let $\phi_{s,\tau}(x,\omega), s,\tau \in T,x \in R^d$ be a continous $R^d$-valued
random field defined on $(\Omega,{\cal F},P)$. Then, for almost all $\omega \in \Omega$,
$\phi_{s,\tau}(\omega)\equiv \phi_{s,\tau}(.,\omega)$ defines a continuous map
from $R^d$ into itself, for any $s,\tau$. Then, let us assume that $\phi_{s,\tau}$ satisfies the
conditions: (i) $\phi_{s,u}(\omega) = \phi_{\tau,u}(\omega)\phi_{s,\tau}(\omega)$,
holds for all $s,\tau,u$, where in the r.h.s. of (i) we have the composition of maps,
(ii) $\phi_{s.s}(\omega) = id$,
for all $s$, where $id$ denotes the identity map,
(iii) $\phi_{s,\tau}(\omega): R^d \to R^d$,
is an onto homeomorphism for all $s,\tau$, and
(iv) $\phi_{s,\tau}(x,\omega)$, is an onto homeomorphism with respect to $x$ 
for all $s,\tau$, and the derivatives are continuous in $(s,\tau,x)$.

Let $\phi_{s,\tau}(\omega)^{-1}$ be the inverse map of $\phi_{s,\tau}(\omega)$.
Then i) and ii) imply that $\phi_{\tau,s}(\omega) = \phi_{s,\tau}(\omega)^{-1}$.
This fact and condition iii) show that $\phi_{s,\tau}(\omega)^{-1}$
is also continuous in $(s,\tau,x)$ and further condition iv) implies that $\phi_{s,\tau}(\omega)^{-1}(x)$
is $k$-times differentiable with respect to $x$. Hence $\phi_{s,\tau}(\omega): R^d \to R^d$
is actually a $C^k$-diffeomorphism of $M$, for all $s,\tau$, the so-called
random diffeomorphic flow. We can regard $\phi_{s,\tau}(\omega)^{-1}(x)$ as a
random field with parameter $(s,\tau,x)$ which is often denoted as $\phi_{s,\tau}^{-1}(x,\omega)$.
Therefore,
\begin{eqnarray}
\phi_{s,\tau}^{-1}(x) = \phi_{\tau,s}(x)
\end{eqnarray}
holds for all $s,\tau,x$ a.s.When we choose the initial time $s$ of $\phi_{s,\tau}(x,\omega)$
to be $0$, we shall write $\phi_\tau (x,\omega)$.

\section*{Acknowledgements}{The auther would like to express his gratitude to the Isaac Newton Institute and the organizers of the 
Geometry and Topology of Fluid Flows Seminar, August-December 2000, for their kind invitation to participate at the Seminar. Partial results of these project were presented at the Seventh International Workshop on Instabilities and Nonequilibrium Structures, Valparaiso, Chile, December 1997, after an invitation by Prof. E. Tirapegui, at the Workshop on Evolution Equations, Univ. of Lisbon, May 1999, after a kind invitation due to
Prof. Elworthy, at the Annual Meeting of the Israel Mathematical Society, Haifa,
May 1999, at the Maths. Department, Univ. of Pisa, February 2000, after an invitation by Prof. F.Flandoli, at the Year 2000 International Conference on
Differential Equations and Dynamical Systems, Kennesaw State Univ., May 2000,
Atlanta. We express our gratitude to all these institutions and conference organizers.

\end{document}